# Using the chi-square test to compare asteroid and laboratory reflectance spectra

Latika Joshi[1], Ines Belkhodja[1], Livneh Naaman[1], Thomas H. Burbine[1], Brian Burt[2]

## Abstract

This study uses the Pearson's chi-square ($\chi^2$) test to analyze the VNIR reflectance spectra of seven asteroids and look for spectral matches among approximately 11,000 laboratory spectra of meteoritic, terrestrial, synthetic, Apollo, and Luna samples. First, we use the $\chi^2$ method to analyze three well-studied asteroids - (4) Vesta, (6) Hebe, and (19) Fortuna - to establish the technique's reliability by attempting to confirm previously predicted spectral matches. The $\chi^2$ test is then be applied to four other asteroids: the Mars trojans (5261) Eureka, (101429) 1998 $VF_{31}$, (311999) 2007 $NS_2$, and (385250) 2001 $DH_{47}$. This study focuses on Mars trojans because of their undetermined origin and possible relationship to Mars. For asteroids that may have undergone space weathering, reddening effects are removed from the spectra to allow for a more accurate $\chi^2$ analysis. The top $\chi^2$ matches among the Mars trojans reveal significant spectral similarities with Martian meteorites and minerals found on Mars, supporting the hypothesis of a Martian origin for the analyzed Mars trojans. However, the possibility that these Mars trojans could be fragments of a disrupted differentiated body or bodies cannot be ruled out. We find that using the $\chi^2$ test with a large number of laboratory spectra is a very useful initial technique for identifying spectral analogs to asteroids.

[1] Mount Holyoke College, South Hadley, MA

[2] Lowell Observatory, Flagstaff, AZ

For queries, please contact Latika Joshi at joshi23l@mtholyoke.edu

## Introduction

Remote sensing is the primary method of estimating the mineralogies of bodies in our Solar System. Planetary bodies (e.g., asteroids, Moon, Mars) are typically observed in the visible and near-infrared because many minerals found on these bodies have diagnostic absorption features in these wavelength regions. However, any analysis of these spectra is typically complicated since mineral bands often overlap or can be suppressed due to the presence of opaques. Many spectral analysis techniques such as radiative transfer modeling or Modified Gaussian modeling (MGM) only work successfully for an initial rough estimate of the possible mineralogy of the surface. At present, there is no standard technique to initially analyze an asteroid spectrum. We hypothesize that the Pearson's chi-square ($\chi^2$) (Pearson 1900; Franke et al. 2012) test is one of the best techniques to first "look" for spectral analogs to an asteroid spectrum.

The $\chi^2$ test is a useful mathematical tool for quantitatively determining spectral similarities between samples measured in laboratories and planetary reflectance spectra taken by a telescope in an unbiased manner (e.g., Popescu et al., 2012; DeMeo et al. 2022; Dibb et al. 2023). Determining $\chi^2$ values involves the following: measuring the variance between observed and expected values (e.g., normalized reflectance spectra), squaring the differences, normalizing by the expected values, and summing said values to obtain a $\chi^2$ value. This method is relatively easy to implement and works best with a large number of laboratory spectra using a variety of compositions. The use of an automated technique allows for assemblages that are possible but not normally considered to be tested as possible spectral analogs.

One possible complication in comparing asteroid and meteorite spectra is that the surface of an asteroid may be affected by space weathering and may not spectrally resemble meteorites that originate from that body or have similar mineralogies. Space weathering typically reddens and darkens the reflectance spectra of non-carbonaceous silicate material (e.g., Yamada et al. 1999; Sasaki et al. 2001) while carbonaceous material tends to have their spectra become more blueish and darker (e.g., Lantz et al. 2018; Thompson et al. 2020). Space weathering is due to the presence of nanophase particles that tend to form due to micrometeorite impacts or solar wind irradiation. The nanophase particles range in composition from metallic Fe-dominated (Noguchi et al. 2011) to FeNi-sulfides (Thompson et al. 2020).

The $\chi^2$ technique's efficacy for the initial comparison of asteroid and laboratory spectra will be tested using a control group of asteroids with well-justified, or "known", mineralogies such as (4) Vesta, (6) Hebe, and (19) Fortuna. Vesta has been linked with the HED (howardite, eucrite, and diogenite) meteorites (e.g., McCord et al. 1970; Larson & Fink 1975), Hebe with the H chondrites (e.g., Gaffey and Gilbert 1998) , and Fortuna with the CM2 chondrites (e.g., Burbine 1998). Testing asteroids with well-grounded compositions, will allow us to determine the veracity of the technique because the $\chi^2$ analysis should yield results in line with each asteroid's known mineralogies. Upon verification, the technique will then be applied to asteroids with much more debated mineralogies ("unknowns").

The "unknowns" will be four Mars Trojans [(5261) Eureka, (101429) 1998 $VF_{31,}$ (311999) 2007 $NS_2$ , and (385250) 2001 $DH_{47}$], Mars trojans have been postulated by a number of researchers to be captured asteroids (e.g., Rivkin et al. 2003, 2007) and/or fragments of Mars (e.g., Borisov et al. 2017; Polishook et al. 2017; Hyodo & Genda 2018). Mars trojans have stable orbits that fall 60 degrees ahead (L4 point) or behind Mars (L5 point) (e.g., Scholl et al. 2005). Only sixteen Mars trojans have currently been identified (Minor Planet Center 2024).

This paper will explore how well the $\chi^2$ method works for the initial analysis of asteroid spectra using laboratory spectra as our comparison dataset. Our goal is to see how well this method works in identifying spectral analogs to asteroids.

## Data

The asteroid reflectance spectra were obtained from a variety of sources. We used only IRTF (Infrared Telescope Facility) SpeX near-infrared spectra taken on Mauna Kea since these spectra have very high resolution. The spectra of main-belt asteroids (4) Vesta, (6) Hebe, and (19) Fortuna and Mars Trojan (5261) Eureka were obtained from the Small Main-Belt Asteroid Spectroscopic Survey (SMASS) website (2024) and are a combination of Kitt Peak CCD (Bus and Binzel 2002) and SpeX spectra. Any overlapping CCD points were deleted from each spectrum. The Eureka visible spectrum was previously discussed in Rivkin et al. (2003) and its visible and near-infrared spectrum was discussed in Rivkin et al. (2007).

Spectra of Mars Trojan (101429) 1998 $VF_{31}$ are from Rivkin et al. (2003, 2007). These spectra are a combination of a Baade 6.5-m telescope (Magellan) CCD and SpeX spectra. Spectra of Mars Trojan (311999) 2007 $NS_2$ and (385250) 2001 $DH_{47}$ are from Polishook et al. (2017) and were just taken with SpeX. The spectrum of 2007 $NS_2$ was smoothed using a running window of width of 0.075 µm. Reduction techniques for the CCD spectra are discussed in Bus and Binzel (2002), Rivkin et al. (2003), and Polishook et al. (2017) and for the IRTF spectra are discussed in Demeo et al. (2009), Rivkin et al. (2007), and Polishook et al. (2017).

**Table 1.** Observing circumstances for asteroids analyzed in this study.

| Asteroid | Type of Spectral Observations | Date of Near-Infrared Observations (day/month/year) | References |
|---|---|---|---|
| (4) Vesta | CCD + SpeX | 18/11/2009 | Bus and Binzel (2002); SMASS (2024) |
| (6) Hebe | CCD + SpeX | 15/01/2008 | Bus and Binzel (2002); SMASS (2024) |
| (19) Fortuna | CCD + SpeX | 29/01/2006 | Bus and Binzel (2002); SMASS (2024) |
| (5261) Eureka | CCD + SpeX | 19/05/2005 | Bus and Binzel (2002); Rivkin et al. (2003, 2007); SMASS (2024) |
| (101429) 1998 $VF_{31}$ | CCD +SpeX | 10/05/2005 | Bus and Binzel (2002); Rivkin et al. (2003, 2007) |
| (311999) 2007 $NS_2$ | SpeX | 13/02/2016 | Polishook et al. (2017) |
| (385250) 2001 $DH_{47}$ | SpeX | 14/02/2016 | Polishook et al. (2017) |

Each asteroid spectrum was compared to 10,884 laboratory reflectance spectra archived in the PDS Geosciences Node Spectral Library. These spectra were taken at the NASA Reflectance Experiment Laboratory (RELAB) at Brown University. The wavelength coverage of the RELAB spectra were ~0.3 to ~2.6 µm. The spectra were taken from a wide variety of samples, which included meteoritic, terrestrial, synthetic, Apollo, and Luna material. The goal was to compare the asteroid spectra to the spectra of as many meteorites, rocks, mineral, and lunar sample spectra as possible to identify any possible spectral analogs.

## Analysis

The $\chi^2$ method compares the observed spectral data from an asteroid with the spectral reflectance data from a variety of samples, and produces a $\chi^2$ value that signifies the degree of similarity between the objects. The lower the $\chi^2$, the closer the spectral match. This quantitative approach provides a systematic and objective method of assessing spectral similarities between asteroids and a variety of laboratory samples by allowing for the consideration of multiple spectral features simultaneously.

DeMeo et al.(2022) implemented a $\chi^2$ matching code, which they used to compare asteroid and meteorite spectra. A simplified version of this code was adapted and modified. The modified code reads in an asteroid spectral file and normalizes the reflectance to unity at either 0.55 µm or 1.215 µm, if the spectrum does not cover the 0.55 µm wavelength. The data are then smoothed using Savitsky-Golay filtering. The higher-resolution laboratory spectra had to be interpolated to the same wavelength range as each asteroid spectrum. The code was adapted to read in a set of spectral files from the PDS Geosciences Node Spectral Library and to compare each spectrum to all of the RELAB spectra. A $\chi^2$ value was calculated for each laboratory spectrum. To confirm the veracity of the $\chi^2$ matching technique, we first analyzed the reflectance spectra of asteroids whose surface compositions are thought to be known.

To mimic the effect of space weathering, a linear slope was divided out from the asteroid spectrum and the resultant spectrum compared to the laboratory spectra. Spectra with removed slopes are listed as ‘de-reddened.” No correction was made for differences in temperature, which can subtly affect absorption features or for phase angle, which can cause reddening at high phase angles. RELAB spectra were measured at room temperature while the asteroids tend to have colder surface temperatures. Visual albedo is only used to see if the asteroid and the matching material have similar visible reflectances.

Each table lists the top twenty $\chi^2$ matches with the laboratory spectra for each asteroid spectrum. DeMeo et al. (2009) previously chose the top twenty $\chi^2$ spectral matches to analyze since they argued that twenty matches is large enough to be representative but small enough to be manually inspected. We list the specimen name, the $\chi^2$ value for its match, specimen type (terrestrial, meteorite, synthetic, Apollo sample, Luna sample, Martian meteorite), meteorite type (if applicable), grain size, and RELAB ID are given. For Apollo and Luna samples, the type of material (e.g., highland; mare basalt) is listed with the specimen name. Meteorite classifications are from the Meteoritical Bulletin Database (2024). Sometimes the accompanying information for each spectrum was not fully descriptive of the sample, so further details were added when available. The top four spectral matches are plotted versus each asteroid spectrum .

## Results

### Control Sample Asteroids

#### (4) Vesta

Asteroid Vesta has long been known to have spectral characteristics similar to basaltic achondritic HED meteorites (e.g., McCord et al. 1970; Larson & Fink 1975). HEDs are predominately composed of varying mixtures of low-Ca pyroxene and plagioclase feldspar and have distinct absorption bands centered at ~0.9 (Band I) and ~1.9 µm (Band II) due to electronic transitions in low-Ca pyroxenes. Pyroxenes with ~0.9 and ~1.9 µm features are called Type B pyroxenes. Vesta's visual albedo (0.43) (Tedesco et al. 2004) is consistent with the higher visible reflectances at ~0.55 µm for HEDs (Ostrowski & Byron 2019). Visual geometric albedos of asteroids and reflectances at ~0.55 µm measured in a laboratory are generally assumed to be directly comparable (Ostrowski & Byron 2019).

The best $\chi^2$ match to Vesta is howardite EET 87513 (**Table 2** and **Figure 1**). EET 87513 is paired with howardite EET 87503 (the fifth best spectral match), which Hiroi et al. (1994) identified as the HED most spectrally similar to Vesta. Visually, EET 87503 is a good spectral match to Vesta with both objects having characteristic absorption bands centered at ~0.9 and ~1.9 µm due to low-Ca pyroxenes with similar band shapes and strengths. The second-best match is with a terrestrial gabbronorite. Gabbronorites are composed of low-Ca pyroxene and plagioclase feldspar, which is a similar mineralogy to eucrites and howardites. The third, fourth, and fifth-best spectral matches are with howardites. Of the top twenty $\chi^2$ matches, 55% are with HED meteorites, reinforcing the hypothesis that these meteorites share a common parent body with Vesta. The spectral similarities between terrestrial rocks, L6 chondrules, a lunar rock, and shergottites with Vesta can be attributed to all these samples having compositions dominated by low-Ca pyroxenes.

**Table 2.** Top twenty $\chi^2$ matches for (4) Vesta organized from lowest to highest $\chi^2$.

| Number | Specimen Name | Chi-Square | Specimen Type | Meteorite Type | Grain Size | RELAB ID |
|---|---|---|---|---|---|---|
| 1 | EET 87513,105 | 0.47 | Meteorite | howardite | <25 µm | camp74 |
| 2 | Gabbronorite | 0.68 | Terrestrial | | <500 µm | c1sw40 |
| 3 | PRA 04402,21 | 0.70 | Meteorite | howardite | <45 µm | c1rm175b |
| 4 | SCO 06040,27 | 0.83 | Meteorite | howardite | <45 µm | c1rm177b |
| 5 | EET 87503,97 | 0.85 | Meteorite | howardite | <25 µm | cgmb68 |
| 6 | Bronzite | 0.85 | Terrestrial | | <45 µm | c1jb236 |
| 7 | GRO 95633,22 | 0.86 | Meteorite | howardite | <45 µm | c1rm187b |
| 8 | Rangala chondrules | 0.95 | Meteorite | L6 | <125 µm | c1dp10 |
| 9 | 76255,75 (Spot E in the breccia matrix) | 1.04 | Apollo sample | | Thin Section | cels24 |
| 10 | QUE 97002,29 | 1.07 | Meteorite | eucrite-pmict | <75 µm | c1mt213a |
| 11 | Olivine Gabbro (weathered surface) | 1.08 | Terrestrial | | <500 µm | c1sw38 |
| 12 | PRA 04401,7 | 1.14 | Meteorite | howardite | <75 µm | c1mt254a |
| 13 | Kapoeta | 1.15 | Meteorite | howardite | <25 µm | camp53 |
| 14 | EETA79001,596 (Lithology A Dark Glass) | 1.15 | Martian Meteorite | shergottite | <50 µm | c1dd59 |
| 15 | Ferric Orthopyroxene | 1.20 | Terrestrial | | <45 µm | c1dd131 |
| 16 | QUE 97002,42 | 1.22 | Meteorite | eucrite-pmict | <45 µm | c1rm206b |
| 17 | EET 87542,42 | 1.23 | Meteorite | eucrite-br | <125 µm | c1rm207a |
| 18 | Orthopyroxene | 1.24 | Terrestrial | | <500 µm | ccsb60 |
| 19 | Orthopyroxene | 1.34 | Terrestrial | | <45 µm | c1op13a |
| 20 | Ferric Orthopyroxene | 1.34 | Terrestrial | | <45 µm | c1dd143 |

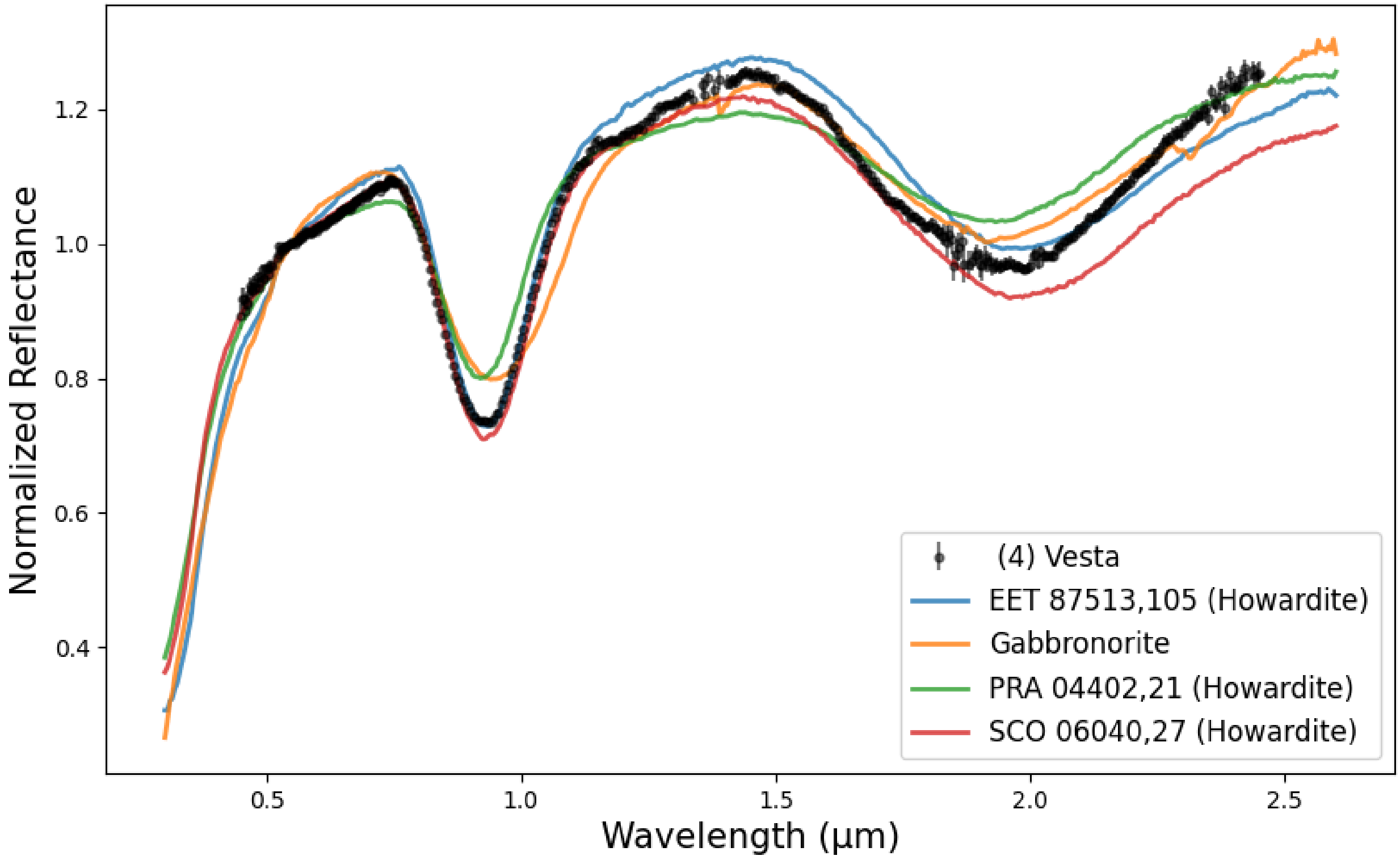

**Figure 1.** Plot of the reflectance spectrum of (4) Vesta (Bus and Binzel 2002; SMASS 2024) versus the spectra of the top four $\chi^2$ spectral matches. All of the spectra are normalized to unity at 0.55 µm. The error bars for the asteroid spectrum are one sigma.

The results for Vesta shows that our $\chi^2$ matching method is valid for determining spectral matches with an asteroid with distinctive spectral features. This result aligns with the widely supported view that Vesta is the source of HED meteorites, as seen by their similar spectral features. This result is confirmation that our $\chi^2$ program will find the best spectral matches to an asteroid.

**(6) Hebe**

Asteroid Hebe has been linked with H chondrites (Gaffey and Gilbert 1996) due to both objects having reflectance spectra with similar absorption band positions in the visible and near-infrared due to low-Ca pyroxene and olivine. H chondrites are composed of low-Ca pyroxene, olivine, and metallic iron plus a number of opaque minerals. Hebe's spectrum is spectrally reddened relative to H chondrites. Hebe's visual albedo (0.27) (Tedesco et al. 2004) is consistent with the higher visible reflectances at ~0.55 µm for H chondrites (Ostrowski & Byron 2019).

Our best $\chi^2$ match with Hebe is with a Lake Hoare sediment from Antarctica (**Table 3** and **Figure 2**) that contains silicates and carbonates (Bishop et al. 2001). The $\chi^2$ matches for asteroid Hebe include a mix of terrestrial samples, meteorites, and synthetic materials. A number of spectra had extremely low $\chi^2$ values, which can be seen in by the extremely good spectral matches in **Figure 2**. However, only one of the top ten matches is with a meteorite, LL4 chondrite Hamlet.

**Table 3 .** Top twenty $\chi^2$ matches for (6) Hebe organized from lowest to highest $\chi^2$. The asteroid spectrum used for the matching has not been corrected for space weathering.

| Number | Specimen Name | Chi-Square | Specimen Type | Meteorite Type | Grain Size | RELAB ID |
|---|---|---|---|---|---|---|
| 1 | Lake Hoare Silicate Carbonate Sediment (core E-5) | 0.19 | Terrestrial | | <125 µm | c1jb194 |
| 2 | Hamlet | 0.21 | Meteorite | LL4 | <125 µm | c1oc02b |
| 3 | Dry Valleys Lake Sediment | 0.24 | Terrestrial | | <125 µm | c1jb672 |
| 4 | Dry Valleys Lake Sediment | 0.25 | Terrestrial | | <125 µm | c1jb655 |
| 5 | Lake Hoare Silicate Carbonate Sediment (core E-4) | 0.27 | Terrestrial | | <125 µm | c1jb193 |
| 6 | Lake Hoare Silicate Carbonate Sediment (core E-10) | 0.30 | Terrestrial | | <125 µm | c1jb204 |
| 7 | Dry Valleys Lake Sediment | 0.32 | Terrestrial | | <125 µm | c1jb652 |
| 8 | Dry Valleys Lake Sediment | 0.32 | Terrestrial | | <125 µm | c1jb653 |
| 9 | Antarctic Lake Sediment | 0.34 | Terrestrial | | <125 µm | c1jb113 |
| 10 | Wöhlerite | 0.36 | Terrestrial | | <45 µm | capp93 |
| 11 | Pinnaroo (metal) | 0.40 | Meteorite | mesosiderite-A4 | chip | c1tb152 |
| 12 | Dry Valleys Lake | 0.41 | Terrestrial | | <125 µm | c1jb670 |
| | Sediment | | | | | |
| 13 | Cynthiana (irradiated with laser) | 0.42 | Meteorite | L/LL4 | <125 µm | c1oc15d15 |
| 14 | Antarctic Lake Sediment | 0.45 | Terrestrial | | <125 µm | c1jb115 |
| 15 | Dry Valleys Lake Sediment | 0.45 | Terrestrial | | <125 µm | C1jb668 |
| 16 | Mixture of 95% enstatite from aubrite Peña Blanca Spring and 5% oldhamite from aubrite Norton County | 0.45 | Meteorite | aubrite | <125 µm | c1tb61 |
| 17 | Pigeonite | 0.48 | Terrestrial | | <45 µm | c1dd162 |
| 18 | Don Quixote Pond Sediment | 0.48 | Terrestrial | | <150 µm | c1jbb51 |
| 19 | TIL 82403,34 | 0.49 | Meteorite | Eucrite-br | chip | cbrm211 |
| 20 | Dry Valleys Lake Sediment | 0.50 | Terrestrial | | <125 µm | c1jb651 |

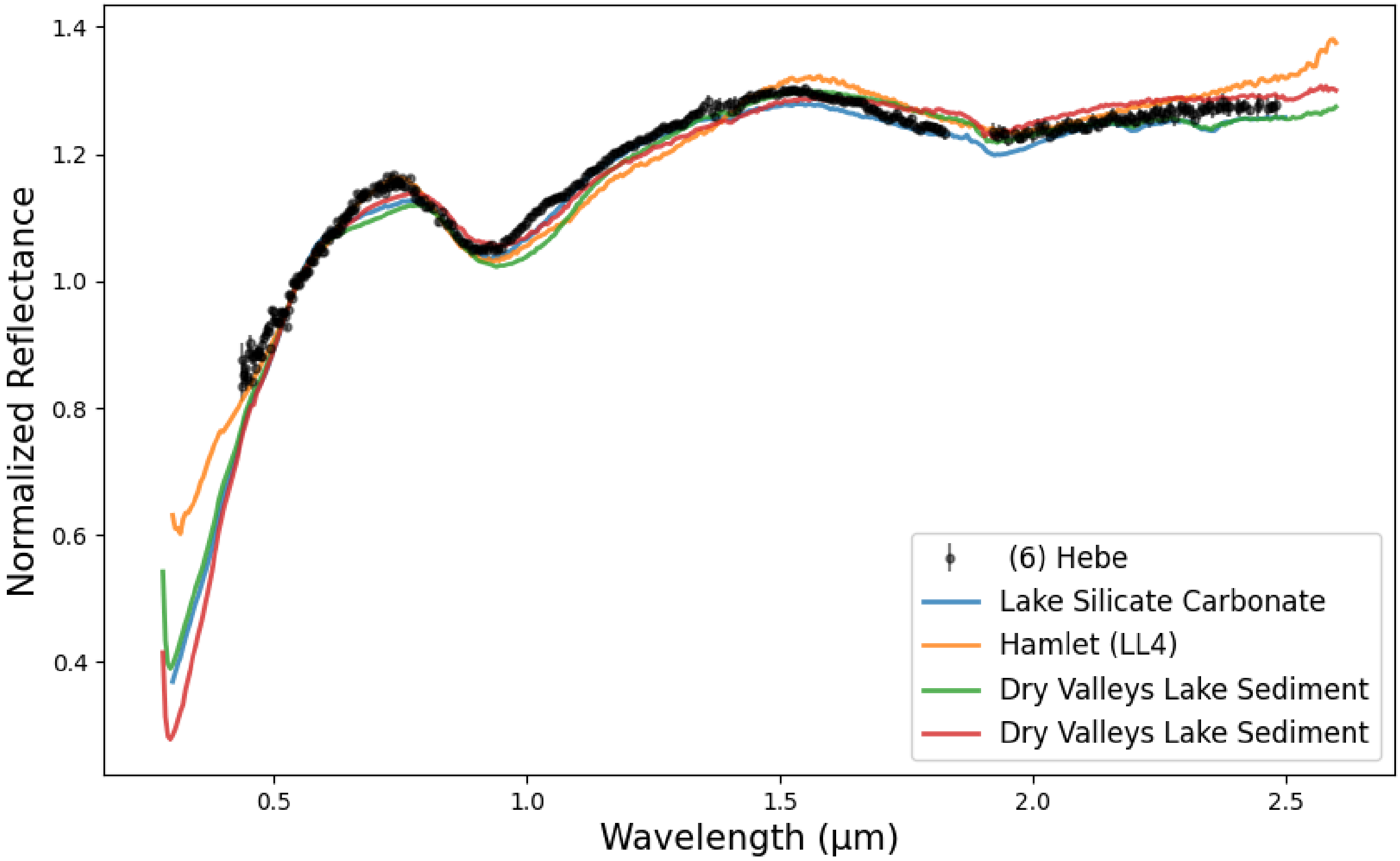

**Figure 2.** Plot of the reflectance spectrum of (6) Hebe (Bus and Binzel 2002; SMASS 2024) versus the spectra of the top four $\chi^2$ spectral matches. All of the spectra are normalized to unity at 0.55 µm. The error bars for the asteroid spectrum are one sigma. The asteroid spectrum has not been corrected for space weathering.

However, the lack of low $\chi^2$ meteorite matches for Hebe argues that the Hebe spectrum should be de-reddened before comparing to the laboratory spectra. The asymmetric structure of the ~0.9 µm band argues for significant olivine abundances on its surface and, as mentioned earlier, olivine is known to space weather at a relatively fast rate. We divide out a linear slope from the Hebe spectrum and then compare that spectrum to the laboratory spectra. We list the best twenty $\chi^2$ matches (**Table 4**) and plot the best four $\chi^2$ matches are plotted in **Figure 5**. A larger number of the matches are with meteorites such as H chondrites, which is consistent with Hebe's interpreted mineralogy; however, our rather simplistic de-reddening can be seen to cause a decrease in spectral slope for Hebe from ~1.5 µm (**Figure 5**). This blue nature of the de-reddened spectrum past ~1.5 µm is probably not characteristic of the original material and is probably leading to larger $\chi^2$ values for the matches.

**Table 4.** Top twenty $\chi^2$ matches for (6) Hebe (de-reddened) organized from lowest to highest $\chi^2$.

| Number | Specimen Name | Chi-Square | Specimen Type | Meteorite Type | Grain Size | RELAB ID |
|---|---|---|---|---|---|---|
| 1 | Basaltic Glass | 0.36 | Terrestrial | | 45-75 µm | cebe254 |
| 2 | Monroe | 0.39 | Meteorite | H4 | 20-250 µm | c1mh07 |
| 3 | Burnwell (irradiated) | 0.40 | Meteorite | H4-an | chip | c1oc21a20 |
| 4 | 62231,14 (highland) | 0.40 | Apollo sample | | bulk | log |
| 5 | 62231,14 (highland) | 0.41 | Apollo sample | | bulk | svls30 |
| 6 | Basaltic Glass | 0.46 | Terrestrial | | 45-75 µm | ccbe254 |
| 7 | Basaltic Glass | 0.50 | Terrestrial | | 45-75 µm | cnbe254 |
| 8 | Fayetteville | 0.51 | Meteorite | H4 | 25-250 µm | cdmb07 |
| 9 | Pulsora | 0.51 | Meteorite | H5 | <150 µm | c1tb120 |
| 10 | Basaltic Glass | 0.52 | Terrestrial | | 45-75 µm | ckbe254 |
| 11 | Basaltic Glass | 0.53 | Terrestrial | | 45-75 µm | cmbe254 |
| 12 | Basaltic Glass | 0.53 | Terrestrial | | <25 µm | c1be70 |
| 13 | DaG 430 | 0.54 | Meteorite | C3-ung | <125 µm | c1mp127 |
| 14 | Hedjaz (irradiated) | 0.54 | Meteorite | L3.7-6 | chip | c1oc16a40 |
| 15 | Cavour | 0.56 | Meteorite | H6 | slab | c1mh24 |
| 16 | Feldspar | 0.56 | Terrestrial | | 45-150 μm | c1sr45a |
| 17 | Basaltic Glass | 0.57 | Terrestrial | | 45-75 μm | clbe254 |
| 18 | Halloysite + Mica + Kaolinite + Quartz | 0.57 | Terrestrial | | <100 μm | c1ev12 |
| 19 | Aragonite | 0.58 | Terrestrial | | <45 μm | cacb44 |
| 20 | Basalt | 0.60 | Terrestrial | | <45 μm | c1fb05 |

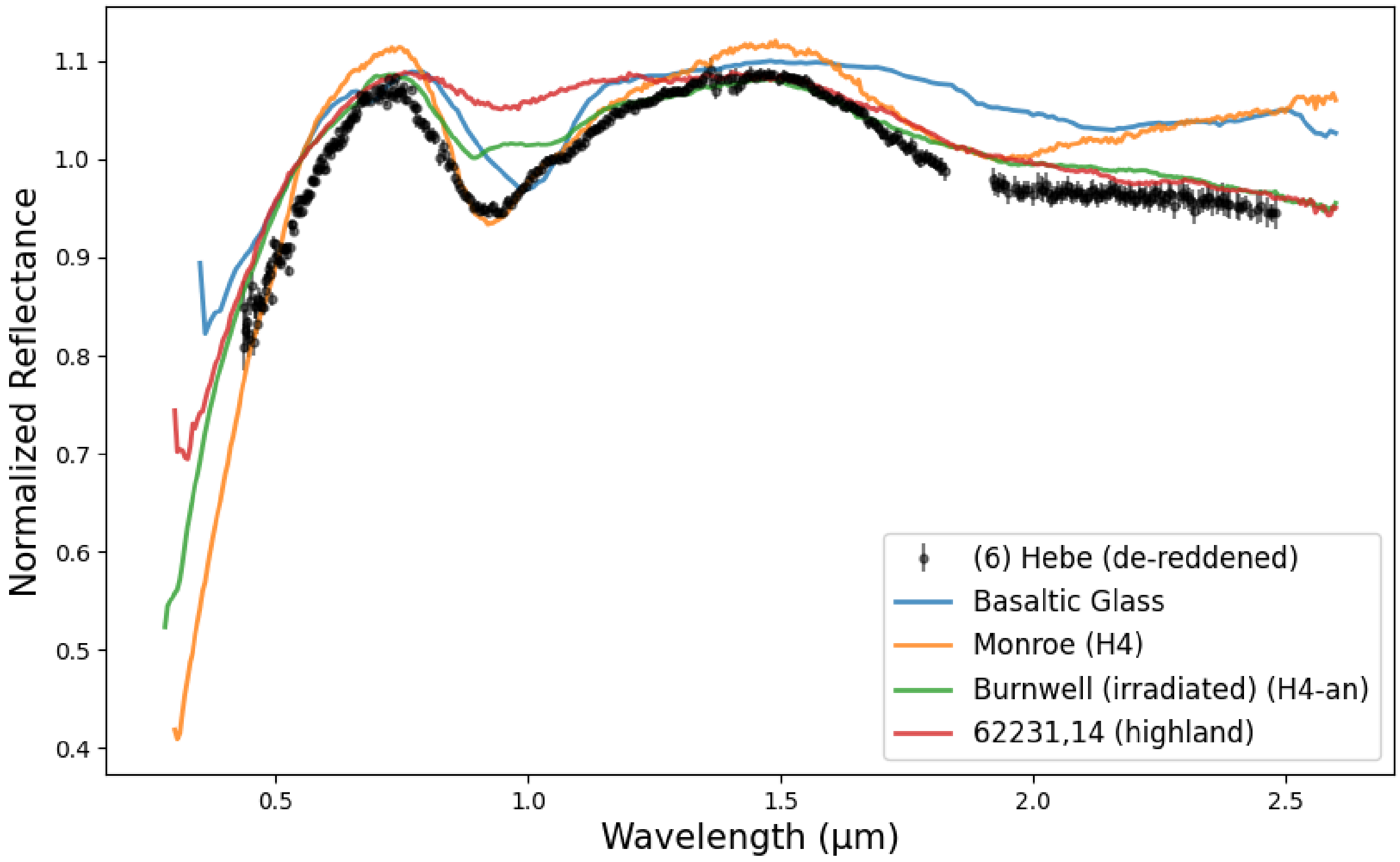

**Figure 3.** Plot of the reflectance spectrum of (6) Hebe (de-reddened) versus the spectra of the top four $\chi^2$ spectral matches. All of the spectra are normalized to unity at 0.55 µm. The error bars for the asteroid spectrum are one sigma.

**(19) Fortuna**

Asteroid Fortuna, has been linked with CM2 chondrites (Burbine 1998) due to the presence of a 0.7 µm band, which tends to be found in CM2 chondrite spectra (e.g., Cloutis et al. 2011) and similarities in spectral slopes. CM2 chondrites are rich in organic compounds and water-bearing minerals. Fortuna's low visual geometric albedo of ~0.06 (Mainzer et al. 2019) is consistent with the visible reflectances at ~0.55 µm for CM2 chondrites (Ostrowski & Byron 2019).

The majority (twelve) of the top twenty $\chi^2$ matches (**Table 5**) for Fortuna are with CM2 chondrites. The top spectral match for asteroid Fortuna is a carbonaceous chondrite, LEW 90500, with an extremely low $\chi^2$, which is indicative of a strong similarity to the asteroid's spectrum. This similarity can be seen in **Figure 4**. Burbine (1998) previously identified LEW 90500 as spectrally similar to Fortuna. The fourth best spectral match is also with a CM2 chondrite. As with Vesta, Fortuna matches very well with meteorites that were predicted to be spectrally analogous with. This result is also confirmation that our $\chi^2$ program will find the best spectral matches to an asteroid.

**Table 5.** Top twenty $\chi^2$ matches for (19) Fortuna organized from lowest to highest $\chi^2$.

| Number | Specimen Name | Chi-Square | Specimen Type | Meteorite Type | Grain Size | RELAB ID |
|---|---|---|---|---|---|---|
| 1 | LEW 90500 | 0.16 | Meteorite | CM2 | <100 µm | c1mb54 |
| 2 | Kerite | 0.20 | Terrestrial | | 100-200 µm | cbms20 |
| 3 | LAR 04316 | 0.26 | Meteorite | aubrite | <45 µm | c1mt331 |
| 4 | MAC 88176,16 | 0.29 | Meteorite | CM2 | <125 µm | c1mp47 |
| 5 | Nogoya | 0.30 | Meteorite | CM2 | <63 µm | c3mb62 |
| 6 | Cold Bokkeveld | 0.31 | Meteorite | CM2 | <63 µm | c1mb61b |
| 7 | EET 83250,7 | 0.34 | Meteorite | CM2 | <125 µm | c2mp40 |
| 8 | Diopside (shocked) | 0.35 | Terrestrial | | <45 µm | c1dd70 |
| 9 | Nuevo Mercurio | 0.36 | Meteorite | H5 | <50 µm | camh53 |
| 10 | Murray | 0.37 | Meteorite | CM2 | <35 µm | c1mt189 |
| 11 | Silica Gel (implanted with ferric nitrate solution) | 0.40 | Synthetic | | 30-70 µm | c2sn106 |
| 12 | Tsarev | 0.40 | Meteorite | L5 | >300 µm | c1ma53 |
| 13 | ALH 85013,32 | 0.43 | Meteorite | CM2 | <125 µm | c1mp23 |
| 14 | EET 90021,10 | 0.43 | Meteorite | CM2 | <125 µm | c1mp43 |
| 15 | Nogoya <35 um | 0.43 | Meteorite | CM2 | <35 µm | c1mt187 |
| 16 | Esquel | 0.44 | Meteorite | pallasite, PMG | slab | c2mb43 |
| 17 | Murchison | 0.52 | Meteorite | CM2 | <90 µm | cbmh52 |
| 18 | Clinochrysotile (heated) | 0.55 | Terrestrial | | bulk | c1cr08 |
| 19 | LON 94101,19 | 0.57 | Meteorite | CM2 | <125 µm | c1mp26 |
| 20 | MET 00630,6 | 0.57 | Meteorite | CM2 | chip | camp225 |

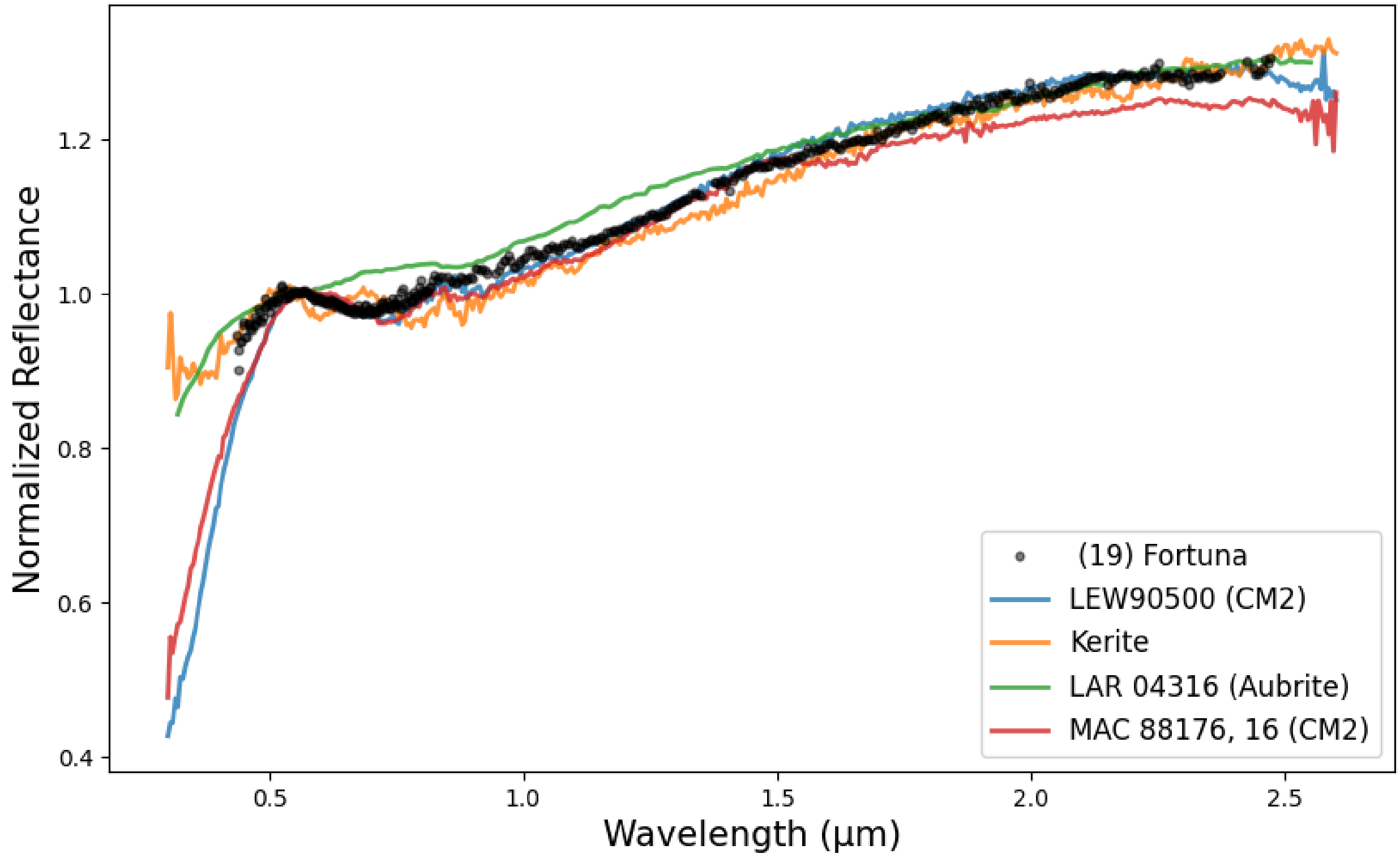

**Figure 4.** Plot of the reflectance spectrum of (19) Fortuna (Bus and Binzel 2002; SMASS 2024) versus the spectra of the top four $\chi^2$ spectral matches. All of the spectra are normalized to unity at 0.55 µm. The error bars for the asteroid spectrum are one sigma.

The second best spectral match is with kerite. Kerite is an organic compound found in terrestrial rocks, whose mineralogies and spectral properties have been used to mimic materials on asteroids with interpreted compositions similar to carbonaceous chondrites (e.g., Kaplan et al. 2018). However, the third best spectral match (aubrite) does not appear to be a good analog for Fortuna. The aubrite spectrum (**Figure 4**) does not have a 0.7 µm band and its reflectance (0.25) at 0.55 µm is much higher than Fortuna's visual albedo (~0.06).

## Asteroids with "unknown" compositions

### (5261) Eureka

Mars Trojan (5261) Eureka is at the L5 point and has been identified as the largest member of a family of asteroids (Christou 2013; Ćuk et al.  2015; Christou et al. 2017). Eureka has a diameter of ~1.3-1.9 km and a geometric visual albedo of ~0.23-0.39 (Trilling et al. 2007; Mainzer et al. 2019). Mars Trojan (5261) Eureka has been linked to angrites due to its reflectance spectrum having such as an asymmetric 1 μm feature and a weak to absent 2 μm feature (Rivkin et al. 2007). Eureka's 1 μm feature also has weak to absent olivine bands, which is consistent with the spectra of angrites (Burbine et al. 2006). Angrites are achondritic meteorites that are composed of a mixture of Al-Ti-bearing diopside-hedenbergite, calcic olivine, and anorthite feldspar (Keil 2012). The diopside-hedenbergite is a Type A pyroxene (e.g., Cloutis & Gaffey 1991; Cloutis 2002), which has an asymmetric 1 μm band and a weak to absent 2 μm band. Spectral modeling by Rivkin et al. (2007) was able to match the reflectance spectrum of Eureka using a mixture of angritic material and a high albedo neutral component.

The $\chi^2$ analysis (**Table 6**) of the Rivkin et al. (2007) Eureka spectrum indicates that Martian meteorite ALHA77005, with a $\chi^2$ value of 1.17, is the closest spectral match (**Figure 5**). ALHA77005 is a shergottite and is composed primarily of olivine (~55 vol%), low-and high-Ca pyroxene (~35 vol%), and maskelynite (~8 vol%) (McSween et al. 1979; Mason 1981). The second and third best matches are with different measurements of the same olivine sample. The fourth and fifth best match is with a synthetic Martian basalt that was altered in water and under high pressure (Cannon et al. 2017). An angrite (Sahara 99555) is the eighth-best spectral match. The range for Eureka's visual geometric albedo is consistent with the visible reflectances at ~0.55 μm for most of these spectra.

The matches for Vesta and Fortuna had much lower $\chi^2$ values than the matches with Eureka and their spectra were much more visually similar to the best-matching laboratory spectra. The reason for the higher $\chi^2$ values for the best-matching laboratory spectra for Eureka appears partially due to the olivine bands in the Eureka spectrum being weaker than the olivine bands in the four best-matching spectra.

**Table 6 .** Top twenty $\chi^2$ matches for (5261) Eureka (Rivkin et al. 2007) organized from lowest to highest. The asteroid spectrum used for the matching has not been corrected for space weathering.

| Number | Specimen Name | Chi-Square | Specimen Type | Meteorite Type | Grain Size | RELAB ID |
|---|---|---|---|---|---|---|
| 1 | ALHA77005 | 1.17 | Martian Meteorite | shergottite | <50 μm | c1dd26 |
| 2 | Olivine ($Fo_{60}$) | 1.59 | Synthetic | | <45 μm | c2dd40 |
| 3 | Olivine ($Fo_{60}$) | 1.71 | Synthetic | | <45 μm | c1dd40 |
| 4 | Synthetic Martian Basalt (altered in water by heating under high pressure) | 1.83 | Synthetic | | 45-75 μm | c1kc48 |
| 5 | Synthetic Martian Basalt (altered in water by heating under high pressure) | 1.90 | Synthetic | | 45-75 μm | c1kc32 |
| 6 | Olivine (heated) | 1.91 | Terrestrial | | <45 μm | c1ol18d |
| 7 | Basaltic Sand | 2.08 | Terrestrial | | <250 μm | c1uh23 |
| 8 | Sahara 99555 | 2.35 | Meteorite | angrite | <125 μm | c1tb57 |
| 9 | Celadonite | 2.38 | Terrestrial | | <45 μm | c1ea20a |
| 10 | Basaltic Ash | 2.49 | Terrestrial | | <400 μm | c1wf14 |
| 11 | Basaltic Sand | 2.57 | Terrestrial | | < 2,000 μm | c1uh05 |
| 12 | EET 87860,14 | 2.70 | Meteorite | CK5/6 | | c1lm12 |
| 13 | Basaltic Soil | 2.74 | Terrestrial | | | c1cw09 |
| 14 | Murchison | 2.75 | Meteorite | CM2 | < 63 μm | cbmb64 |
| 15 | A-881955,70 | 2.78 | Meteorite | CM2 | < 125 μm | c1mp37 |
| 16 | Murchison | 2.90 | Meteorite | CM2 | <123 μm | camb64 |
| 17 | Olivine-rich Basalt (mixed with synthetic glasses) | 2.92 | Synthetic | | 45-75 μm | c1kc52 |
| 18 | Basaltic Tuff | 2.92 | Terrestrial | | <400 μm | c1bu32 |
| 19 | Y-984028,74 | 3.09 | Martian meteorite | shergottite | <45 μm | c1dd118 |
| 20 | Basaltic Tephra | 3.15 | Terrestrial | | <400 μm | c1bu10 |

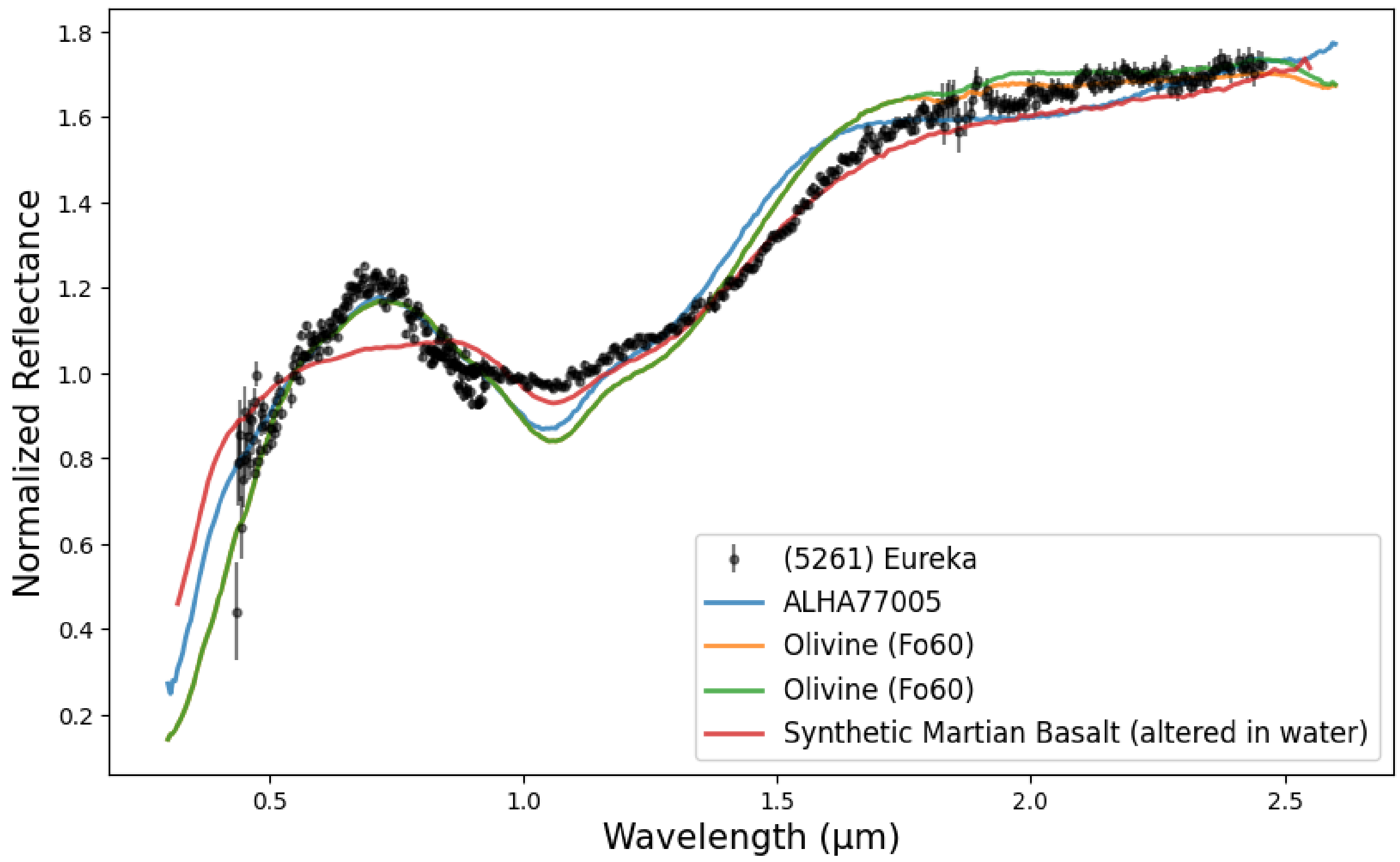

**Figure 5.** Plot of the reflectance spectrum of (5261) Eureka (Bus and Binzel 2002; Rivkin et al. 2007; SMASS 2024) versus the spectra of the top four $\chi^2$ spectral matches. All of the spectra are normalized to unity at 0.55 μm. The error bars for the asteroid spectrum are one sigma. The asteroid spectrum has not been corrected for space weathering.

To see if space weathering could potentially be affecting the number of meteorite matches, we de-reddened the Eureka spectrum. The top twenty $\chi^2$ matches are in **Table 7** and a plot of the top four matches are plotted in **Figure 6**. The top match is with GRA 06129, which is an achondritic meteorite that contains greater than 75% sodium-rich plagioclase plus olivine, low- and high-pyroxenes, phosphates and sulfides (Day et al. 2009). Mikouchi & Miyamoto (2008) found that a thin section of GRA 06129 contained modal abundances of 82% plagioclase, 9% olivine, 6% augite, 1% orthopyroxene. Olivine tends to dominate the spectral properties of olivine-plagioclase mixtures (e.g., Serventi et al. 2013). Augite is a Type A pyroxene and would be expected to contribute to the asymmetric nature of the Band I. Day et al. (2009) interprets GRA 06129 as being from the crust of an incompletely differentiated asteroid. The other three top spectral matches are with olivine and the olivine-dominated CK chondrite LEW 87009 (Schwarz & Mason 1988; Kallemeyn et al. 1991). Other good matches also tend to be for spectra dominated by olivine. The range for Eureka's visual geometric albedo is also consistent with the visible reflectances at ~0.55 μm for most of these spectra.

**Table 7 .** Top 20 $\chi^2$ matches for de-reddened (5261) Eureka organized from lowest to highest.

| Number | Specimen Name | Chi-Square | Specimen Type | Meteorite Type | Grain Size | RELAB ID |
|---|---|---|---|---|---|---|
| 1 | GRA 06129 (pristine dark material) | 0.85 | Meteorite | achondrite-ung | bulk | c1mt261 |
| 2 | Olivine (Fa35) | 0.99 | Synthetic | | <45μm | c1dd90p |
| 3 | LEW 87009,4 | 1.16 | Meteorite | CK6 | bulk | c1lm17 |
| 4 | LEW 87009,4 | 1.16 | Meteorite | CK6 | <63μm | c1mp05 |
| 5 | Olivine (Fa30) | 1.23 | Synthetic | | <45μm | c1dd89p |
| 6 | Mercury surface analog | 1.49 | Synthetic | | 2-25 μm | c1xx01 |
| 7 | Plagioclase (93 wt%) - Olivine (7 wt%) | 1.51 | Terrestrial | | 45-75 μm | c1mx72c4 |
| 8 | Plagioclase (95 wt%) - Olivine (5 wt%) | 1.55 | Terrestrial | | 45-75 μm | c1mx71c4 |
| 9 | Plagioclase (90 wt%) - Olivine (10 wt%) | 1.64 | Terrestrial | | 45-75 μm | c1mx73c4 |
| 10 | Volcanic Hawaiite Glass | 1.70 | Synthetic | | <45 μm | c1rm72 |
| 11 | Plagioclase (93 wt%) - olivine (7 wt%) | 1.72 | Terrestrial | | 0-1000 μm | c1mx72g |
| 12 | Gabbronorite | 1.82 | Terrestrial | | 0-500 μm | c1sw27 |
| 13 | Plagioclase (93 wt%) - Olivine (7 wt%) | 1.91 | Terrestrial | | 0-1000 μm | c1mx72h |
| 14 | Plagioclase (98 wt%) - Olivine (2 wt%) | 1.93 | Terrestrial | | 45-75 μm | c1mx105c |
| 15 | Irradiated Olivine | 1.93 | Terrestrial | | Thin Section | rkpo88 |
| 16 | LEW87009,28 | 1.95 | Meteorite | CK6 | 0-125 μm | c1mb88 |
| 17 | Almahata Sitta | 2.00 | Meteorite | ureilite-an | 125-500µm | c1mt354b |
| 18 | Selasvann Dolomite | 2.00 | Terrestrial | | 125-250µm | c1jbe61e |
| 19 | Olivine (Fa30) | 2.02 | Synthetic | | <45µm | c1dd116p |
| 20 | Dolomite | 2.02 | Terrestrial | | 45-90µm | cbcb03 |

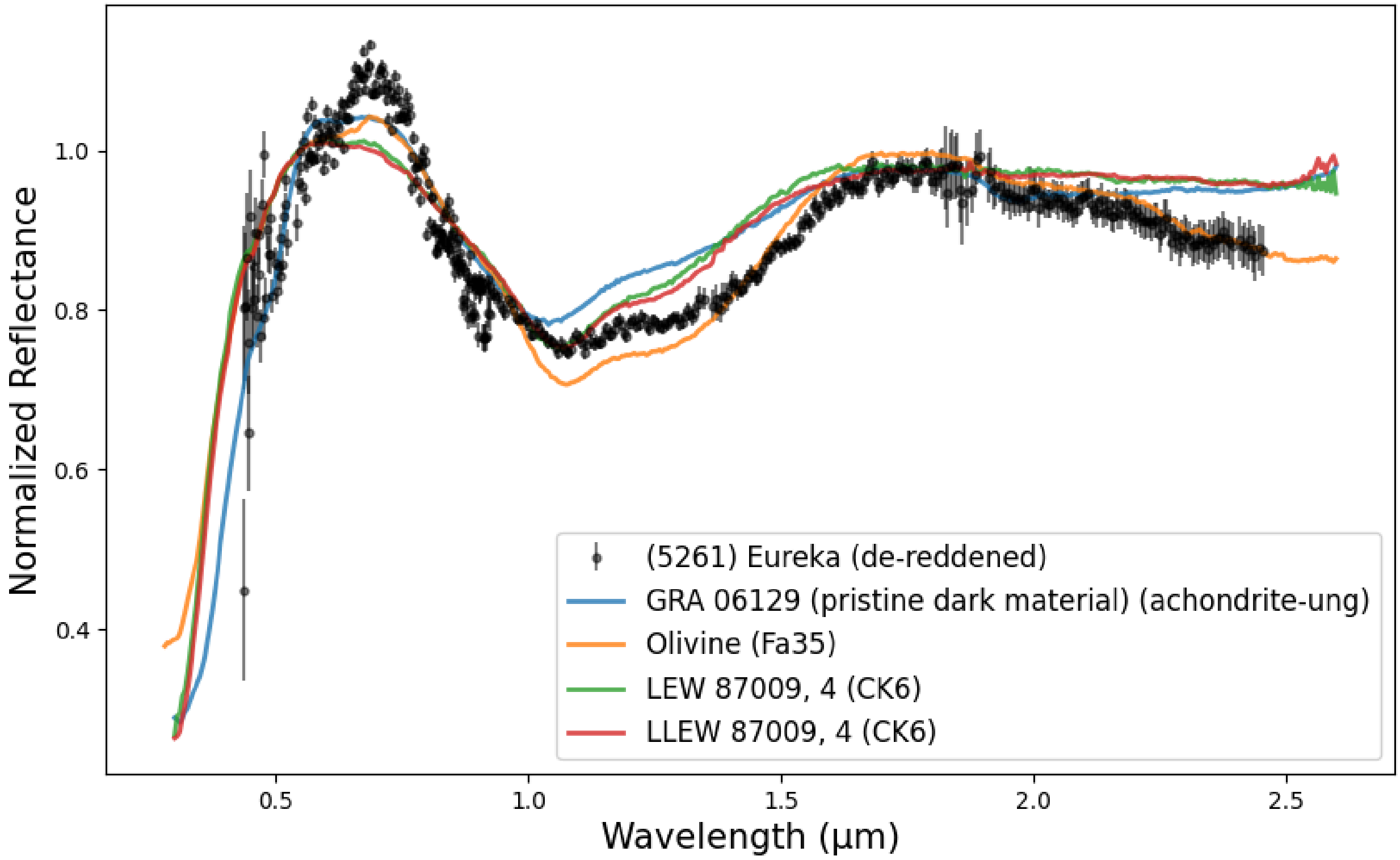

**Figure 6.** Plot of the reflectance spectrum of (5261) Eureka (de-reddened) versus the spectra of the top four $\chi^2$ spectral matches. All of the spectra are normalized to unity at 0.55 µm. The error bars for the asteroid spectrum are one sigma.

## (101429) 1998 $VF_{31}$

Mars Trojan (101429) 1998 $VF_{31}$ is found at the L5 point but has not been identified as part of the Eureka family due to having different orbital elements (Christou 2013). The diameter of 1998 $VF_{31}$ is 0.8 km (Trilling et al. 2007). Rivkin et al. (2007) found that 1998 $VF_{31}$ had a visible and near-infrared reflectance spectrum typical of S-type asteroids with spectral features dominated by low-Ca pyroxenes. Spectral modeling by Rivkin et al. (2007) found that the spectrum of 1998 VF$_{31}$ was best matched by a mixture of metallic iron and primitive achondritic material. Christou et al. (2021) obtained new reflectance spectra of 1998 VF$_{31}$ and noted spectral similarities between this body and a variety of pyroxene-dominated material such as primitive achondrites, Martian crust, and lunar samples through curve matching.

Our top twenty $\chi^2$ matches for 1998 VF$_{31}$ are listed in **Table 8** and the top four spectral matches are plotted in **Figure 7**. The lowest $\chi^2$ values are extremely high (>7) compared to the best matches for Vesta, Hebe, Fortuna, and Eureka and this bigger discrepancy between the asteroid and laboratory spectra can be seen in **Figure 7**. The best matches are with samples dominated by low-Ca pyroxene (Type B) absorption bands; however, the laboratory pyroxene absorption bands tend to be much stronger than those found in the asteroid spectrum. The fourth-best match is with the lodranite NWA 7674, which is a primitive achondrite (Ruzicka et al. 2015) and contains minerals such as olivine, low- and high-Ca pyroxene, and metallic iron. There are also a number of top spectral matches with lunar samples.

**Table 8.** Top twenty $\chi^2$ matches for (101429) 1998 VF$_{31}$ (Bus and Binzel 2002; Rivkin et al. 2007) organized from lowest to highest. The asteroid spectrum used for the matching has not been corrected for space weathering

| Number | Specimen Name | Chi-Square | Specimen Type | Meteorite Type | Grain Size | RELAB ID |
|---|---|---|---|---|---|---|
| 1 | Orthopyroxene (UV irradiated) | 7.28 | Terrestrial | | pellet | c1kk40pu |
| 2 | Enstatite (heated) | 7.79 | Terrestrial | | <45 µm | c1op14c |
| 3 | Orthopyroxene (heated) | 8.21 | Terrestrial | | <45 µm | c1op13d |
| 4 | NWA 7674 (treated with ETG) | 8.41 | Meteorite | lodranite | <125 µm | c1mt310t |
| 5 | 72,255,338 (impact melt) | 8.54 | Apollo sample | | bulk | cals99 |
| 6 | Enstatite | 8.54 | Terrestrial | | <30 µm | c1dh05 |
| 7 | Enstatite (heated) | 8.60 | Terrestrial | | <45 µm | c1op14d |
| 8 | 68,501,601 | 8.60 | Apollo | | >125 µm | c1lr44 |
| 9 | Luna 20 soil 2014,6 (highland) | 8.63 | Luna sample | | 94-250µm | cclu02 |
| 10 | Olivine + Orthopyroxene + Metal (ratio of 30:40:30) | 8.66 | Terrestrial | | 45-90μm | c2sc07 |
| 11 | 62231,58 (highland) | 8.71 | Apollo sample | | 125-250μm | kdls140 |
| 12 | 60019,9004,82 (regolith breccia) | 8.74 | Apollo sample | | 75-250μm | cdls17 |
| 13 | 62231.58 (highland) | 8.76 | Apollo sample | | 125-250μm | kcls140 |
| 14 | 72,255,338 (impact melt) | 8.78 | Apollo sample | | chip | c1ls99 |
| 15 | 61,221,160 (highland) | 8.87 | Apollo sample | | 20-45μm | c1lr106 |
| 16 | Enstatite (heated) | 8.88 | Terrestrial | | <45μm | c1op14b |
| 17 | Pyroxene + Olivine + Plagioclase + Ilmenite | 8.90 | Terrestrial | | 25-75 μm | c2xs09 |
| 18 | Landes (silicate inclusion) | 8.91 | Meteorite | iron, IAB-MG | <125 μm | c1tb60 |
| 19 | Château-Renard (laser irradiated) | 8.99 | Meteorite | L6 | <125 μm | c1oc11d35 |
| 20 | 62231.58 (highland) | 9.00 | Apollo sample | | 125-250 μm | cals140 |

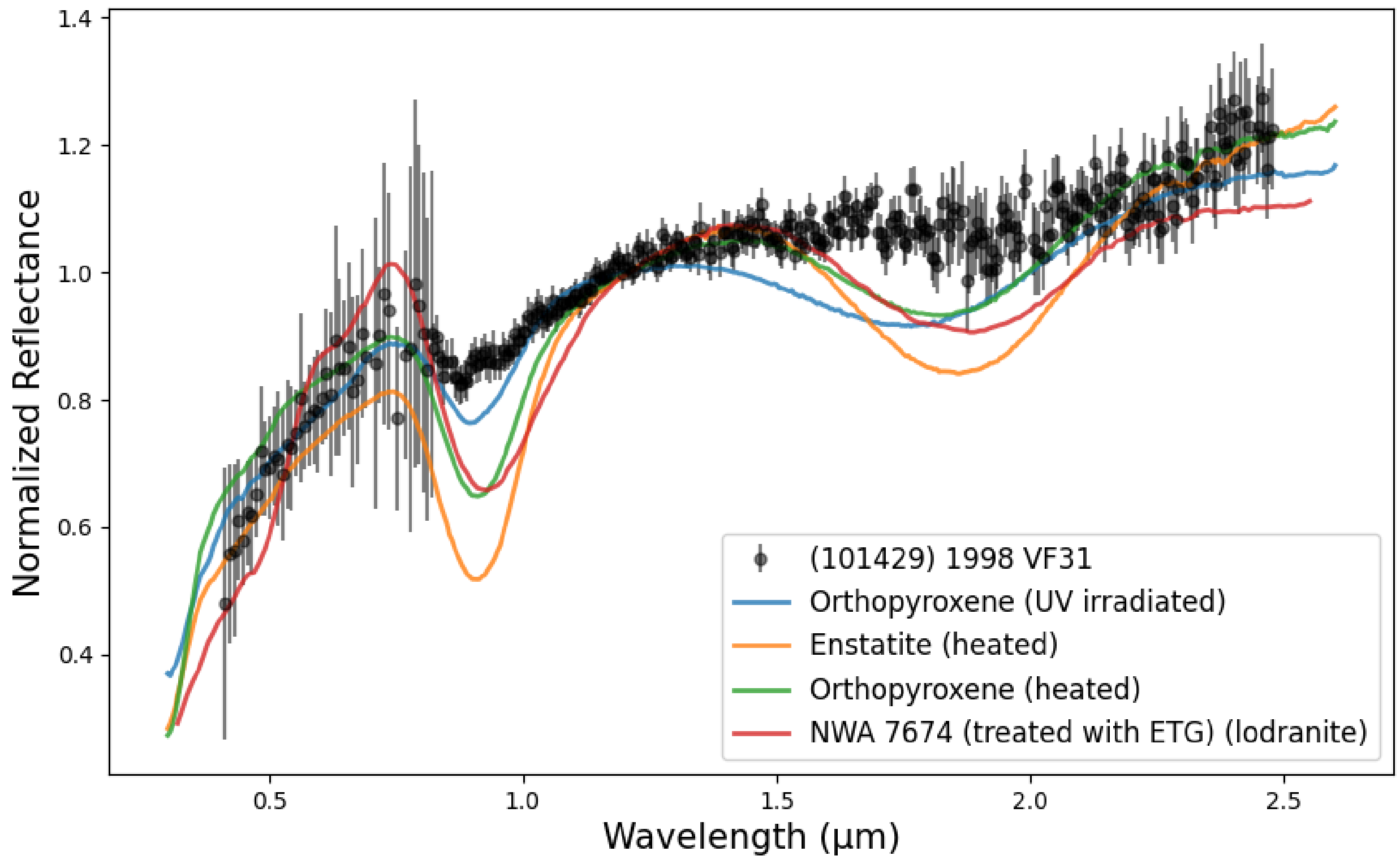

**Figure 7.** Plot of the reflectance spectrum of (101429) 1998 $VF_{31}$ (Bus and Binzel 2002; Rivkin et al. 2007) versus the spectra of the top four $\chi^2$ spectral matches. All of the spectra are normalized to unity at 1.215 μm. The error bars for the asteroid spectrum are one sigma. The asteroid spectrum has not been corrected for space weathering.

To see if space weathering could potentially be affecting the values for the best $\chi^2$ spectral matches even though 1998 $VF_{31}$ does not appear to have an olivine-dominated surface, we de-reddened the 1998 $VF_{31}$ spectrum. The top twenty $\chi^2$ matches are in **Table 9** and the plot of the top four matches are plotted in **Figure 8**. The $\chi^2$ values are slightly lower but still extremely high (>5). The best matches are still with spectra with low-Ca pyroxene absorption bands. The fourth-best match is with the howardite PRA 04402 (McBride et al. 2007), which contains an approximately equal mixture of eucritic and diogenitic material (Herrin et al. 2011).

**Table 9.** Top twenty $\chi^2$ matches for (101429) 1998 $VF_{31}$ (de-reddened) organized from lowest to highest.

| Number | Specimen Name | Chi-Square | Specimen Type | Meteorite Type | Grain Size | RELAB ID |
|---|---|---|---|---|---|---|
| 1 | Enstatite | 5.37 | Terrestrial | | <45 μm | c1dd167 |
| 2 | Enstatite | 5.40 | Terrestrial | | <45 μm | c1jb238 |
| 3 | Bronzite | 5.47 | Terrestrial | | <25 μm | c1jb479 |
| 4 | PRA 04402,21 | 5.65 | Meteorite | howardite | <125 μm | c1rm175a |
| 5 | Enstatite | 5.69 | Terrestrial | | <45 μm | capp107 |
| 6 | Ourique | 5.71 | Meteorite | H4 | <150 μm | c1mt07 |
| 7 | Harzburgite | 5.76 | Terrestrial | | slab | c1sw03 |
| 8 | Bamble enstatite | 5.76 | Terrestrial | | <45 μm | cgpp107 |
| 9 | Bamble enstatite | 5.77 | Terrestrial | | <45 μm | cfpp107 |
| 10 | PCA 02014,7 | 5.79 | Meteorite | howardite | <40 μm | c1mt240 |
| 11 | PCA 02013,6 | 5.81 | Meteorite | howardite | <40 μm | c1mt239 |
| 12 | Bamble enstatite <45 um | 5.82 | Terrestrial | | <45 μm | cepp107 |
| 13 | Ausson | 5.83 | Meteorite | L5 | <150 μm | c1mt84 |
| 14 | Lake Silicate Carbonate (Lake Hoare)-9 | 5.83 | Terrestrial | | bulk | c1jb198c |
| 15 | EETA 79002,146 | 5.84 | Meteorite | diogenite | <25 μm | cgmb67 |
| 16 | Bamble enstatite | 5.85 | Terrestrial | | <45 μm | cbpp107 |
| 17 | Gabbro | 5.87 | Terrestrial | | 63-125 μm | c1rg51 |
| 18 | Enstatite | 5.90 | Terrestrial | | <45 μm | c1dd166 |
| 19 | Allegan (almost all metal removed) | 5.91 | Meteorite | H5 | bulk | c1tb104 |
| 20 | Chiang Khan (almost all metal removed) | 5.96 | Meteorite | H6 | <150 µm | c1tb132 |

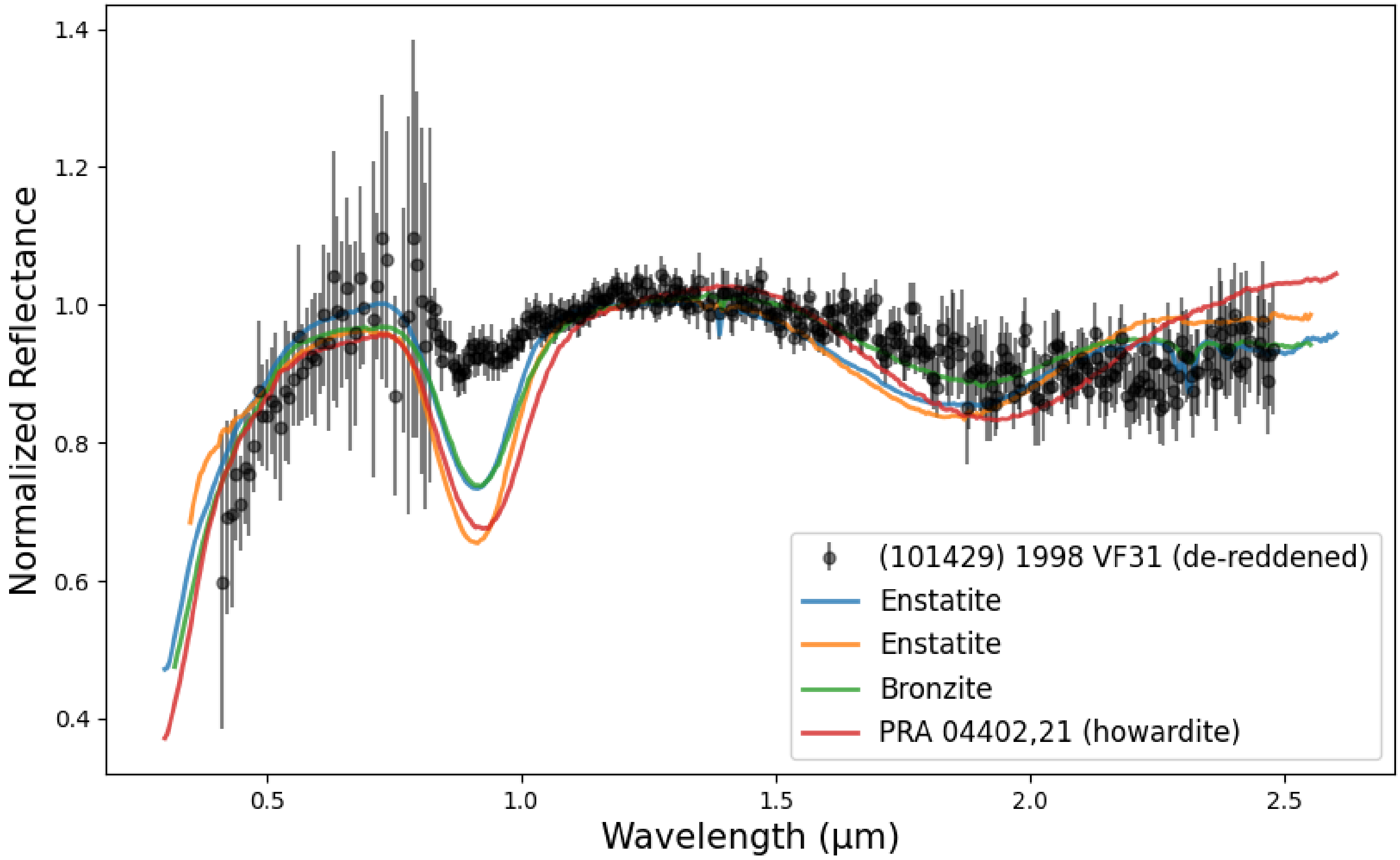

**Figure 8.** Plot of the reflectance spectrum of (101429) 1998 $VF_{31}$ (de-reddened) versus the spectra of the top four $\chi^2$ spectral matches. All of the spectra are normalized to unity at 0.55 µm. The error bars for the asteroid spectrum are one sigma.

## (311999) 2007 $NS_2$

Mars Trojan (311999) 2007 $NS_2$ is part of the Eureka family (Christou 2013). Observations of 2007 $NS_2$ by Borisov et al. (2017) and Polishook et al. (2017) noted that the spectrum had an asymmetric 1 µm feature, consistent with an olivine-dominated surface. Both Borisov et al. (2017) and Polishook et al. (2017) noted that 2007 $NS_2$ had spectral properties similar to Eureka. Polishook et al. (2017) also noted the spectral similarity between 2007 $NS_2$ and the olivine-dominated Martian meteorite Chassigny.

Our top twenty $\chi^2$ matches for 2007 $NS_2$ are listed in **Table 10** and the top four spectral matches are plotted in **Figure 9**. The top twenty $\chi^2$ matches for 2007 $NS_2$ include a range of terrestrial and synthetic samples but no meteoritic samples. Our best match is with a shocked Type A pyroxene. The second and fourth spectral matches are also with high-Ca pyroxenes. The third, fifth, and sixth best spectral matches are with olivine. The 2007 $NS_2$ spectrum, which has been smoothed and is slightly noisy, can be seen to be matched by both high-Ca pyroxene and olivine.

**Table 10.** Top twenty $\chi^2$ matches for (311999) 2007 $NS_2$ organized from lowest to highest. The asteroid spectrum used for the matching has not been corrected for space weathering

| Number | Specimen Name | Chi-Square | Specimen Type | Meteorite Type | Grain Size | RELAB ID |
|---|---|---|---|---|---|---|
| 1 | Pyroxene (shocked) | 1.89 | Terrestrial | | 250-500 µm | cdrs77 |
| 2 | Clinopyroxene (shocked) | 2.53 | Terrestrial | | bulk | ccsr04 |
| 3 | Olivine | 2.98 | Terrestrial | | 45-90 µm | c1ol12 |
| 4 | Diopside | 3.07 | Terrestrial | | 45-150 µm | c1sr42a |
| 5 | Olivine (oxidized) | 3.20 | Terrestrial | | <45 µm | capo05 |
| 6 | Olivine | 3.31 | Synthetic | | <45 µm | c1dd44 |
| 7 | Augite Forsterite Glass | 3.31 | Synthetic | | 38-63 µm | c1ck39 |
| 8 | Altered Lava Phyllosilicate | 3.44 | Terrestrial | | 1-30 µm | c1jm153 |
| 9 | Green olivine (irradiated) | 3.63 | Terrestrial | | 45-75 µm | c1po81cp2 |
| 10 | Labradorite + Olivine | 3.74 | Terrestrial | | 45-75 µm | c1mx110c |
| 11 | Augite + Montmorillonite | 3.81 | Synthetic | | <106 µm | c1er06p |
| 12 | Cronstedtite | 3.89 | Terrestrial | | <45 µm | cacr21 |
| 13 | Pseudotachylyte clasts | 3.92 | Terrestrial | | <115 µm | c1fm02 |
| 14 | San Carlos olivine (irradiated with UV) | 4.00 | Terrestrial | | Pressed Powder Pellet | c1kk39pu |
| 15 | Illite | 4.01 | Terrestrial | | <125 µm | c1jb782 |
| 16 | Metamorphic Rock | 4.08 | Terrestrial | | bulk | c1sr02 |
| 17 | Green olivine (laser irradiated) | 4.23 | Terrestrial | | 45-75 µm | c1po81cp1 |
| 18 | Augite Forsterite Glass | 4.24 | Synthetic | | 38-63 µm | c1ck38 |
| 19 | Hebei olivine | 4.37 | Terrestrial | | 45-75 µm | kayz01 |
| 20 | Augite + Montmorillonite | 4.41 | Synthetic | | <106 μm | c1er08p |

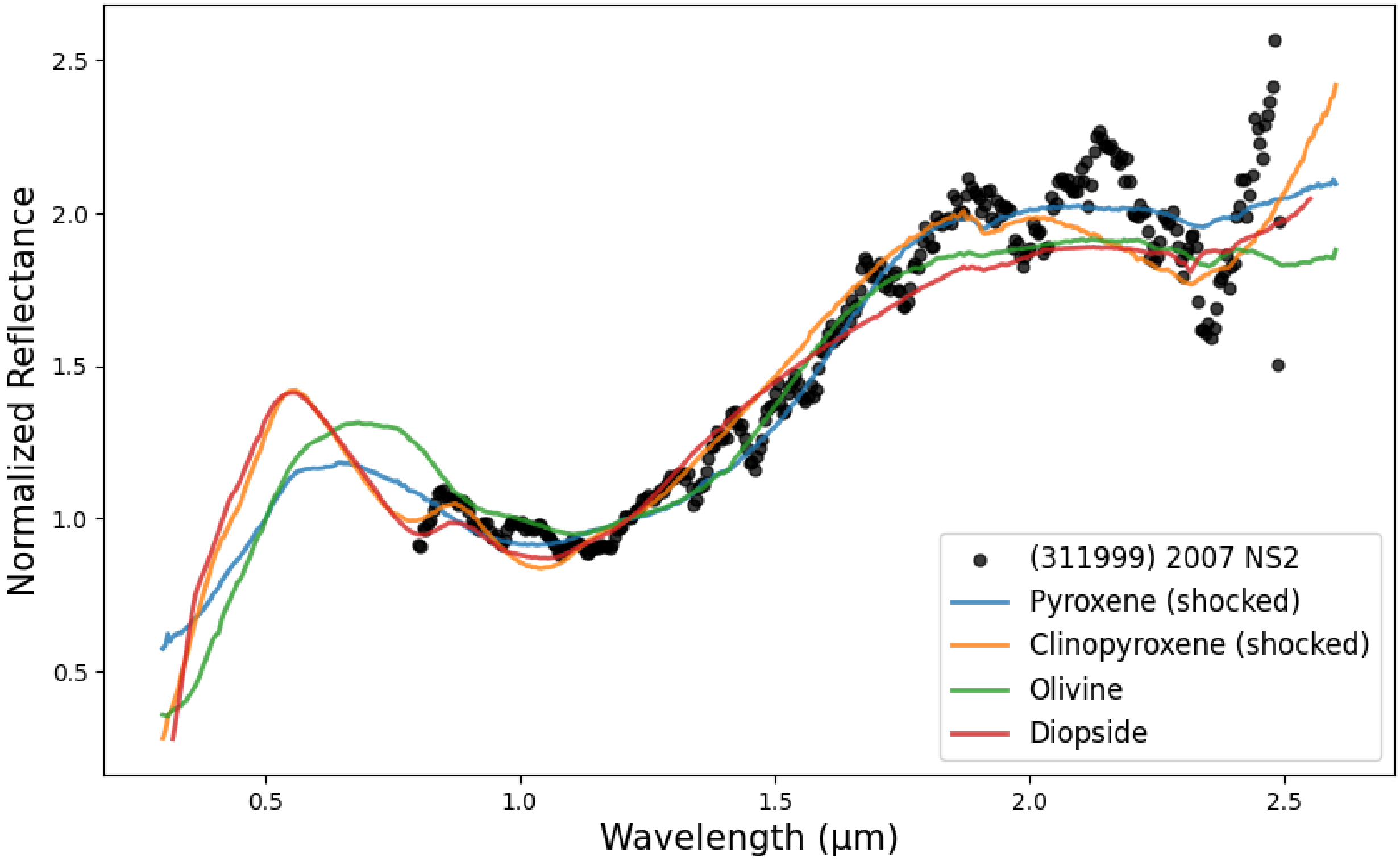

**Figure 9.** Plot of the reflectance spectrum of (311999) 2007 $NS_2$ versus the spectra of the top four $\chi^2$ spectral matches. All of the spectra are normalized to unity at 1.215 μm. The error bars for the asteroid spectrum are not available since the spectrum of 2007 $NS_2$ was smoothed using a running window of width of 0.075 μm. The asteroid spectrum has not been corrected for space weathering.

To see if space weathering could potentially be affecting the number of meteorite matches, we de-reddened the 2007 $NS_2$ spectrum. The top twenty $\chi^2$ matches are in **Table 11** and the plot of the top four matches are plotted in **Figure 10**. There is now one meteorite match (slab of the Imilac pallasite), which is the thirteenth best match, but the rest of the matches are with terrestrial and synthetic samples. A number of the top matches are with carbonates such as dolomite and calcite, which are found on the Martian surface (e.g., Hazen et al. 2023), but not in the abundances necessary (Polomba et al. 2009; Wray et al. 2016) to produce carbonate fragments that are hundreds of meters in diameter like the Mars trojans. Zastrow & Glotch (2021) found carbonate abundances up to only ~35% in the Jezero Crater on Mars using CRISM (Compact Reconnaissance Imaging Spectrometer for Mars) spectra.

**Table 11.** Top twenty $\chi^2$ matches for (311999) 2007 $NS_2$ (de-reddened) organized from lowest to highest $\chi^2$.

| Number | Specimen Name | Chi-Square | Specimen Type | Meteorite Type | Grain Size | RELAB ID |
|---|---|---|---|---|---|---|
| 1 | Weathered Pyroxene | 1.89 | Terrestrial | | crystal | c4sw21 |
| 2 | Varnish | 2.54 | Terrestrial | | slab | cfrd01 |
| 3 | Selasvann Dolomite | 2.65 | Terrestrial | | 90-125 µm | c1jbe61d |
| 4 | Calcite | 2.80 | Terrestrial | | <45 µm | cacb64 |
| 5 | Selasvann Dolomite | 2.88 | Terrestrial | | 250-1000 µm | c1jbe61f |
| 6 | Selasvann Dolomite | 2.93 | Terrestrial | | 75-90 µm | c1jbe61c |
| 7 | Olivine (Fa30) | 2.94 | Synthetic | | <45 µm | c1dd89p |
| 8 | Selasvann Dolomite | 3.09 | Terrestrial | | 125-250 µm | c1jbe61e |
| 9 | Feldspar | 3.14 | Terrestrial | | 125-250 µm | sbpf16 |
| 10 | Augite | 3.17 | Terrestrial | | slab | c1jb478 |
| 11 | Feldspar | 3.17 | Terrestrial | | 125-250 µm | cbpf16 |
| 12 | Olivine (Fa35) | 3.18 | Synthetic | | <45 µm | c1dd90p |
| 13 | Imilac | 3.21 | Meteorite | pallasite, PMG | slab | c8mb41 |
| 14 | Plagioclase | 3.26 | Terrestrial | | Flat surface | csw22a |
| 15 | Plagioclase | 3.35 | Terrestrial | | Flat surface | casw22 |
| 16 | Selasvann Dolomite | 3.37 | Terrestrial | | 45-75 µm | c1jbe61b |
| 17 | Dolomite | 3.49 | Terrestrial | | 45-90 µm | cbcb03 |
| 18 | Rock 4E (California) | 3.53 | Terrestrial | | slab | c3rk15 |
| 19 | Labradorite | 3.56 | Terrestrial | | 150-180 µm | cepl155q2 |
| 20 | Labradorite | 3.57 | Terrestrial | | 150-180 µm | cfpl155q2 |

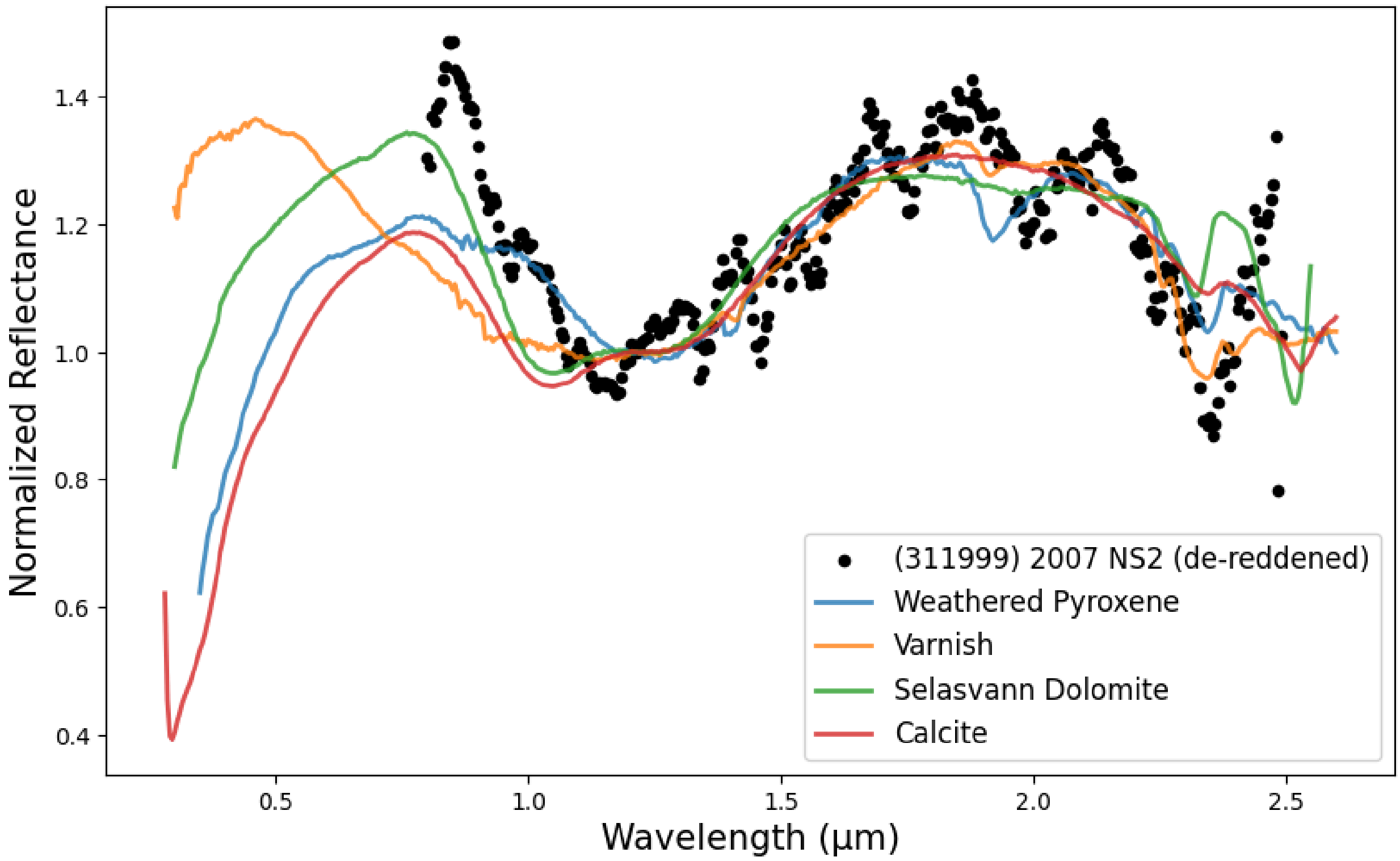

**Figure 10.** Plot of the reflectance spectrum of (311999) 2007 $NS_2$ (de-reddened) versus the spectra of the top four $\chi^2$ spectral matches. All of the spectra are normalized to unity at 1.215 μm. The error bars for the asteroid spectrum are given since the spectrum of 2007 $NS_2$ was smoothed using a running window of width of 0.075 μm.

## (385250) 2001 $DH_{47}$

Mars Trojan (385250) 2001 $DH_{47}$ is part of the Eureka family (Christou et al. 2021). Both Borisov et al. (2017) and Polishook et al. (2017) noted the spectral similarity of 2001 $DH_{47}$ to 2007 $NS_2$. The Polishook et al. (2017) spectrum of 2001 $DH_{47}$ is less noisy than the spectra of 2007 $NS_2$. Borisov et al. (2017) and Polishook et al. (2017) also noted that 2007 $NS_2$ had an asymmetric 1 μm feature consistent with an olivine-dominated surface and spectrally similar to Eureka. Polishook et al. (2017) also identified 2007 $NS_2$ as having similar spectral properties to the olivine-dominated Martian meteorite Chassigny.

The top twenty $\chi^2$ matches are in **Table 12** and the plot of the top four matches are plotted in **Figure 11**. As with 2007 $NS_2$, the top twenty $\chi^2$ matches for 2001 $DH_{47}$ include a range of terrestrial and synthetic samples but no meteoritic samples. Matches tend to be high-Ca Type A pyroxenes and olivine.

**Table 12:** Top twenty $\chi^2$ matches for (385250) 2001 $DH_{47}$ organized from lowest to highest $\chi^2$. The asteroid spectrum used for the matching has not been corrected for space weathering

| Number | Specimen Name | Chi-Square | Specimen Type | Meteorite Type | Grain Size | RELAB ID |
|---|---|---|---|---|---|---|
| 1 | Diopside | 1.44 | Terrestrial | | 45-150 µm | c1sr42a |
| 2 | Pyroxene (shocked) | 1.51 | Terrestrial | | 250-500 µm | cdrs77 |
| 3 | Clinopyroxene (shocked) | 1.53 | Terrestrial | | bulk | ccsr04 |
| 4 | Cronstedtite | 1.84 | Terrestrial | | <45 µm | cacr21 |
| 5 | San Carlos Olivine (UV irradiated) | 2.02 | Terrestrial | | <75 µm | c1kk39pu |
| 6 | Illite | 2.04 | Terrestrial | | <125 µm | c1jb782 |
| 7 | Green Olivine (laser irradiated) | 2.23 | Terrestrial | | 45-75 µm | c1po81cp 2 |
| 8 | Olivine | 2.25 | Terrestrial | | 45-90 µm | c1ol12 |
| 9 | Augite Forsterite Glass | 2.33 | Synthetic | | 38-63µm | c1ck39 |
| 10 | Olivine ($Fo_{20}$) | 2.36 | Synthetic | | <45µm | c1dd44 |
| 11 | Magnetite | 2.41 | Terrestrial | | <45µm | camg16 |
| 12 | Augite Forsterite Glass | 2.44 | Synthetic | | 38-63 µm | c1ck38 |
| 13 | Metamorphic Rock | 2.63 | Terrestrial | | Nonparticula te | c1sr02 |
| 14 | Green Olivine | 2.64 | Terrestrial | | <45 µm | c2dd78 |
| 15 | San Carlos Olivine (Baked & Laser irradiated) | 2.69 | Terrestrial | | N/A | c1dd53 |
| 16 | Illite | 2.71 | Terrestrial | | <45 µm | c2jb262 |
| 17 | Illite | 2.79 | Terrestrial | | 75-89 µm | c1jb782c |
| 18 | Tan Olivine | 2.30 | Terrestrial | | <45 µm | c2dd76 |
| 19 | Plagioclase + Labradorite | 2.94 | Terrestrial | | 45-75 µm | c1mx110c |
| | Olivine | | | | | |
| 20 | Augite + Amorphous Silica | 2.97 | Synthetic | | <106 µm | c1er01 |

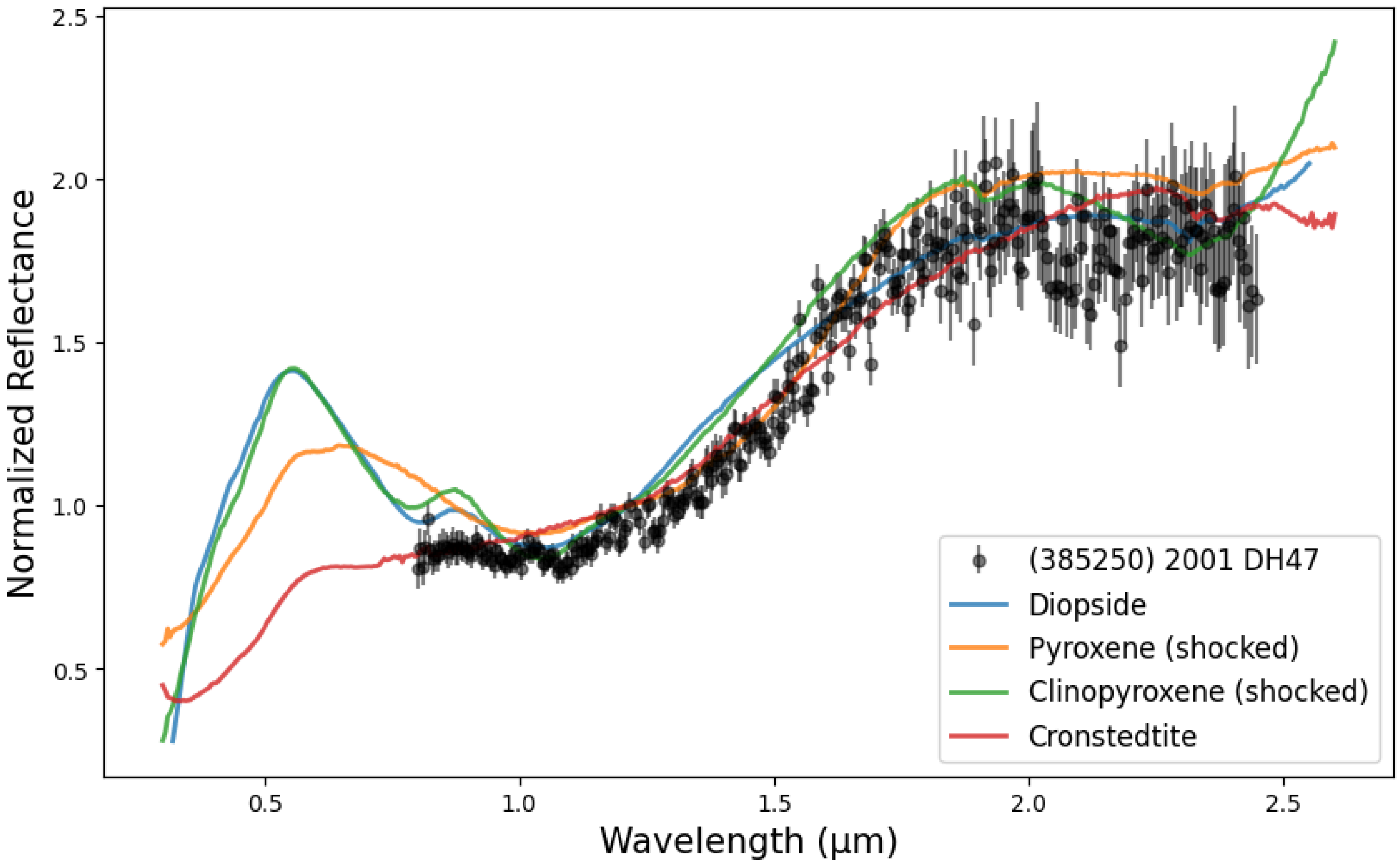

**Figure 11.** Plot of the reflectance spectrum of (385250) 2001 $DH_{47}$ versus the spectra of the top four $\chi^2$ spectral matches. All of the spectra are normalized to unity at 1.215 µm. The error bars for the asteroid spectrum are one sigma. The asteroid spectrum has not been corrected for space weathering.

To see if space weathering could potentially be affecting the number of meteorite matches, we de-reddened the 2007 $NS_2$ spectrum. The top twenty $\chi^2$ matches are in **Table 13** and the plot of the top four matches are plotted in **Figure 12**. There are now two meteorite matches (both are for a slab of an Imilac pallasite) with the rest of the matches being with terrestrial and synthetic samples. As with (311999) 2007 $NS_2$, a number of the top matches are with carbonates such as dolomite and calcite.

**Table 13.** Top twenty $\chi^2$ matches for (385250) 2001 $DH_7$4 (de-reddened) organized from lowest to highest.

| Number | Specimen Name | Chi-Square | Specimen Type | Meteorite Type | Grain Size | RELAB ID |
|---|---|---|---|---|---|---|
| 1 | Rock 4E (California) | 1.21 | Terrestrial | | slab | c3rk15 |
| 2 | Selasvann Dolomite | 1.44 | Terrestrial | | 75-90 μm | c1jbe61c |
| 3 | Imilac | 1.45 | Meteorite | Pallasite, PMG | slab | c8mb41 |
| 4 | Calcite | 1.47 | Terrestrial | | <45 μm | cacb64 |
| 5 | 10 wt% Gypsum + 90 wt% Dolomite | 1.49 | Synthetic | | 90-150 μm | c1jbe72 |
| 6 | Rock 8C (California) | 1.52 | Terrestrial | | slab | c4rk15 |
| 7 | Selasvann Dolomite | 1.56 | Terrestrial | | 45-75 μm | c1jbe61b |
| 8 | Iron Dolomite | 1.68 | Terrestrial | | <125 μm | c1jb779 |
| 9 | Weathered Pyroxene | 1.70 | Terrestrial | | crystal | c4sw21 |
| 10 | Varnish | 1.82 | Terrestrial | | slab | cfrd01 |
| 11 | Imilac | 1.88 | Meteorite | Pallasite, PMG | slab | c2mb41 |
| 12 | Selasvann Dolomite | 1.89 | Terrestrial | | 90-125 μm | c1jbe61d |
| 13 | Olivine (Fa30) | 2.00 | Synthetic | | <45 μm | c1dd89p |
| 14 | Volcanic basalt | 2.02 | Terrestrial | | 250-500 μm | c1rb42b |
| 15 | Augite | 2.21 | Terrestrial | | slab | c1jb478 |
| 16 | Plagioclase | 2.34 | Terrestrial | | slab | casw22 |
| 17 | Plagioclase (sintered) | 2.37 | Synthetic | | 45-125 μm | c1pl142d |
| 18 | Feldspar | 2.38 | Terrestrial | | 63-125 μm | cipf15 |
| 19 | Plagioclase | 2.39 | Terrestrial | | slab | csw22a |
| 20 | 93 wt% Plagioclase + 7 wt% Olivine | 2.41 | Terrestrial | | <1000 μm | c1mx72g |

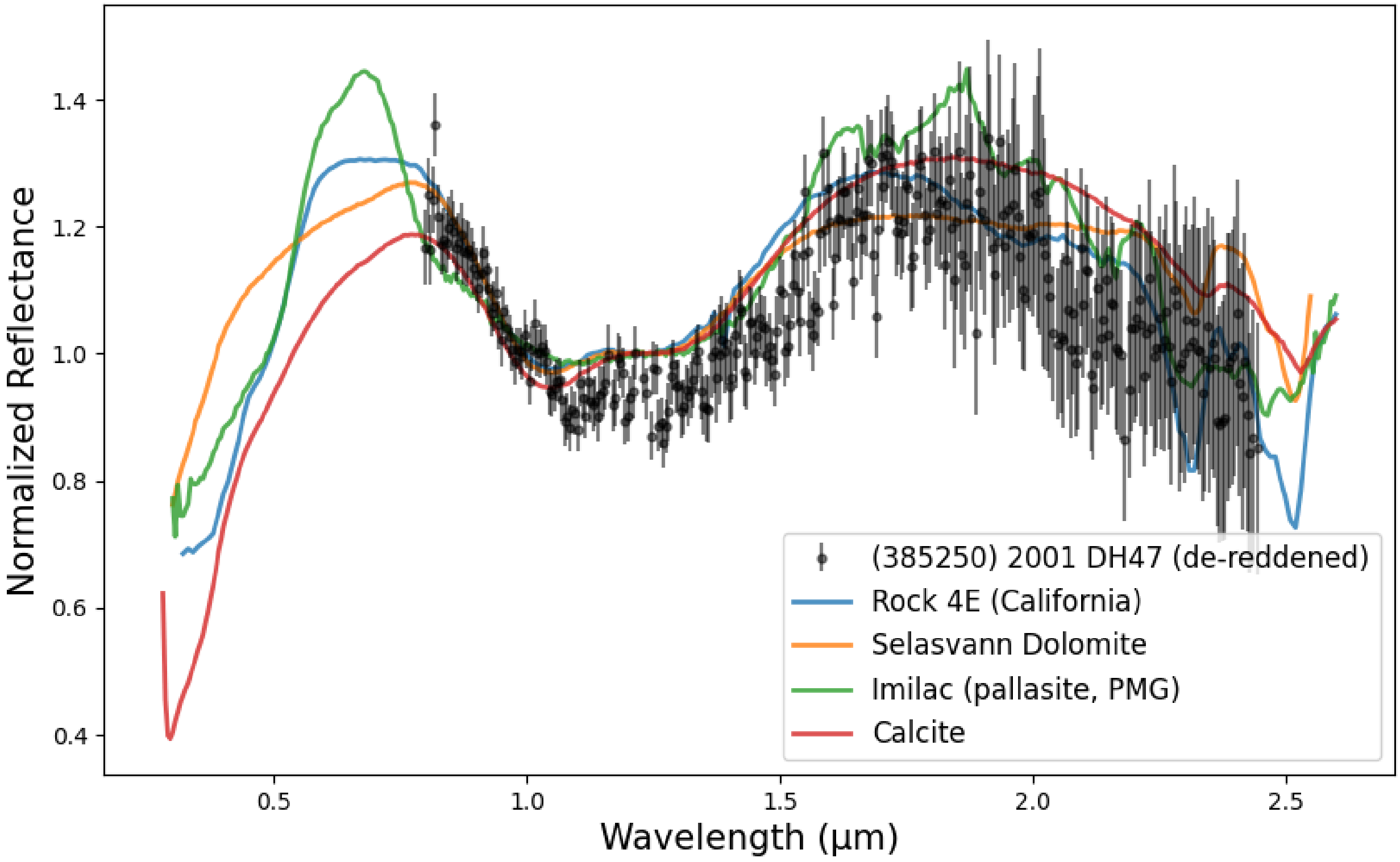

**Figure 12.** Plot of the reflectance spectrum of (385250) 2001 $DH_{47}$ (de-reddened) versus the spectra of the top four $\chi^2$ spectral matches. All of the spectra are normalized to unity at 1.215 μm. The error bars for the asteroid spectrum are one sigma.

## Conclusions

The $\chi^2$ test is an extremely effective way in determining spectral similarities between asteroid and laboratory spectra. We confirmed this method to be extremely reliable as spectral matches previously predicted for asteroids Vesta and Fortuna were found to be some of the best $\chi^2$ matches to these objects among ~11,000 RELAB laboratory spectra of meteoritic, terrestrial, synthetic, Apollo, and Luna samples. For a body that appears to be space-weathered like Hebe, removing a linear slope (de-reddening) increases the number of meteoritic matches. We suggest that all asteroid spectra should be compared to a large number of laboratory spectra using the $\chi^2$ test before any assumptions are made about an asteroid's mineralogy. Other methods, such as machine learning, may also give considerable insight into spectral analogs for asteroids but we suggest that these techniques should be applied to an asteroid spectrum after the $\chi^2$ test.

By comparing an asteroid spectrum to not just meteorite spectra, it is possible to determine possible spectral analogs with material not currently present in our meteorite collections. One stumbling block in using spectra from the RELAB database is that some of the spectra are not labeled with enough mineralogical information to understand what is causing their spectral properties.

We used the $\chi^2$ test to determine the best spectral analogs to a number of Mars trojans [(5261) Eureka, (101429) 1998 $VF_{31}$, (311999) 2007 $NS_2$, and (385250) 2001 $DH_{47}$] observed by a number of researchers (Bus and Binzel 2002; Rivkin et al. 2007; Polishook et al. 2017; SMASS 2024). Because it is unknown if the Mars Trojans surface had been space-weathered, we analyzed the original spectra and ones that were de-reddened. For Eureka, the best $\chi^2$ matches for both the original and the de-reddened spectra are for olivine-dominated assemblages. The best $\chi^2$ match for the original Eureka spectrum is with the Martian meteorite ALHA77005 while the best $\chi^2$ match for the de-reddened Eureka spectrum is with the ungrouped achondrite GRA 06129, which is composed predominantly of plagioclase with much smaller amounts of olivine and augite (Type A pyroxene). The $\chi^2$ values for the best matches to 1998 $VF_{31}$ (original and de-reddened) were very high but the best spectral matches tended to be for assemblages rich in low-Ca pyroxenes. Mars trojans 2007 $NS_2$, and 2001 $DH_{47}$ had similar spectral properties (Borisov et al. 2017; Polishook et al. 2017). We find that the best spectral matches for 2007 $NS_2$ and 2001 $DH_{47}$ include augite, olivine, and carbonates.

The best $\chi^2$ matches to the Mars trojans tended to be minerals commonly found on Mars such as olivine and low- and high-Ca pyroxene (e.g., Hazen et al. 2023). Orbital spectroscopy of Mars has identified minerals such as olivine, pyroxene, and plagioclase on the surface of Mars (e.g., Ehlmann & Edwards 2014). Additionally, the matches included a number of synthetic samples, with some of the material created to mimic the Martian mineralogy, and Martian meteorites. Moreover, many meteorites have mineralogies dominated by olivine or low- and high-Ca pyroxene and some of these meteorites were also part of the best $\chi^2$ matches. The igneous nature of the Martian crust can also be roughly duplicated by igneous processes that occurred on some asteroids.

Fragments of planetary bodies being identified as asteroids appear possible. Near-Earth asteroid (469219) Kamoʻoalewa has been postulated to be a fragment of the Moon due to having a reflectance spectrum (Sharkey et al. 2021) and orbit (Castro-Cisneros et al. 2023; Jiao et al. 2024) consistent with lunar material. The Martian moons Phobos and Deimos have been postulated to contain Martian material (e.g., Hyodo et al. 2017; Glotch et al. 2018).

The significance of a Martian origin for Mars trojans is that these objects may be easier to sample and return to Earth than the surface of Mars due to the planet's atmosphere and significant gravity. Samples from Mars that have not been exposed to Earth's atmosphere have the potential to provide better insight into Mars' geologic history and the potential for life. Sample return from Mars trojans would be expected to either return Martian material or asteroidal material that has never been previously returned to Earth. However, the high inclinations relative to the ecliptic of Mars trojans is a challenge for doing sample return at a "reasonable" cost (Wickhusen et al. 2023). Sample return from Mars trojans is the only way to conclusively know their origin.

## Acknowledgments

LJ and THB would like to thank the Remote, In Situ, and Synchrotron Studies for Science and Exploration 2 (RISE2) Solar System Exploration Research Virtual Institute (SSERVI) (NASA grant 80NSSC19M0215) for support. This research utilizes meteorite reflectance spectra acquired at the NASA RELAB facility at Brown University. The authors would like to thank David Polishook and Andrew Rivkin for supplying their Mars Trojan reflectance spectra. The authors would also like to thank M. Darby Dyar for their helpful comments.